\documentclass[floatfix,showpacs,prd,aps,tightenlines,amsmath,amsfonts,amssymb,nofootinbib]{revtex4}

\usepackage{graphicx}
\usepackage{color}
\usepackage{dcolumn}% Align table columns on decimal point

\newcommand{\bea}{\begin{eqnarray}}
\newcommand{\eea}{\end{eqnarray}}

\begin{document}

\title{Detecting quasinormal modes of binary black hole mergers with
second-generation gravitational-wave detectors}

\author{Takashi Nakamura, Hiroyuki Nakano and Takahiro Tanaka}

\affiliation{Department of Physics, Kyoto University, Kyoto 606-8502, Japan}

\begin{abstract}
Recent population synthesis simulations of Pop III stars suggest that
the event rate of coalescence of $\sim 30M_\odot$--$30M_\odot$ binary black holes
can be high enough for the detection by the second generation gravitational wave detectors.
The frequencies of chirp signal as well as quasinormal modes
are near the best sensitivity of these detectors so that it would be possible
to confirm Einstein's general relativity. Using the WKB method,
we suggest that for the typical value of
spin parameter $a/M\sim 0.7$ from numerical relativity results of
the coalescence of binary black holes,
the strong gravity of the black hole space-time at around the radius $2M$,
which is just $\sim 1.17$ times the event horizon radius,
would be confirmed as predicted by general relativity.
The expected event rate with the signal-to-noise ratio $> 35$ needed
for the determination of the quasinormal mode frequency
with the meaningful accuracy
is $0.17$--$7.2$~${\rm events~yr^{-1}~(SFR_p/(10^{-2.5}~M_\odot~yr^{-1}~Mpc^{-3}))}
\cdot (\rm [f_b/(1+f_b)]/0.33)$
where ${\rm SFR_p}$ and ${\rm f_b}$ are the peak value of the Pop III star
formation rate and the fraction of binaries, respectively.
As for the possible optical counter part, if the merged black hole
of mass $M\sim 60M_\odot$ is in the interstellar matter with
$n\sim 100~{\rm cm^{-3}}$ and the proper motion of black hole
is $\sim 1~{\rm km~s^{-1}}$, the luminosity is $\sim 10^{40}~{\rm erg~ s^{-1}}$
which can be detected up to $\sim 300~{\rm Mpc}$, for example,
by Subaru-HSC and LSST with the limiting magnitude 26. 
\end{abstract}


\maketitle

%%%%%%%%%%%%%%%%%%%%%%%%%%%%%%%%%%%%%%%%
\section{Introduction}
%%%%%%%%%%%%%%%%%%%%%%%%%%%%%%%%%%%%%%%%

The second generation gravitational wave detectors such as
Advanced LIGO (aLIGO)~\cite{TheLIGOScientific:2014jea}, 
Advanced Virgo (AdV)~\cite{TheVirgo:2014hva}, 
and KAGRA~\cite{Somiya:2011np,Aso:2013eba} are now about
to reach the observable mean distance of $\sim 200$~Mpc
for the chirp signal of neutron star (NS)--NS binaries.
Among them, aLIGO is now operating with the range of
$60$--$80$~Mpc from September 18, 2015
to mid-January in 2016~\cite{nov_2015_news}.
One of the most important targets is NS--NS mergers
which might be associated with Short Gamma Ray Bursts (SGRBs).
There are three methods to determine the expected event rate of NS--NS mergers.
The first one is to use the observed NS--NS binaries.
Kim, Kalogera and Lorimer~\cite{Kim:2002uw} performed Monte Carlo simulations
using the already existing pulsar surveys.
Assuming the pulsar distribution function in our galaxy, 
the luminosity function of pulsars and pulsar beaming factors, 
they concluded that the event rate is $17.9^{+21}_{-10.6}~{\rm events~yr^{-1}}$
for aLIGO, AdV and KAGRA adopting their model 1
although there is a factor of 38 difference in the rate among their 27 models.
Kalogera et al.~\cite{Kalogera:2003tn}
corrected the rate since the new double pulsar PSR J0737-3039 was found.
The new rate becomes $186.8^{+470.5}_{-148.7}~{\rm events~yr^{-1}}$
using their model 6 which predicts about three times larger rate than that of model 1.
Kalogera et al.~\cite{Kalogera:2003tn} corrected the error
in their simulations which reduced the rate to
$83.0^{+209.1}_{-66.1}~{\rm events~yr^{-1}}$.
The latest expected event rate by Kim et al.~\cite{Kim:2013tca}
is $8.0^{+10}_{-5}~{\rm events~yr^{-1}}$ at $95\%$ confidence level,
adopting their model 6. This further reduction mainly
comes from the new beaming factor correction obtained from the observations
of PSR J0737-3039.
 
The second method is to use the observed event rate of SGRBs
assuming that the SGRB is NS--NS and/or NS--black hole (BH) mergers.
Here, we should note that no NS--BH binary has been observed
as a pulsar binary.
According to Fong et al.~\cite{Fong:2013eqa},
the number of SGRBs with the confirmed redshift $z$ by {\it Swift} and HETE-2 is
at most 20 or so. However, BATSE on CGRO~\cite{CGRO} found
$\sim 900$~SGRBs between 1990 and 2000 without the information of redshift $z$.
To determine the redshifts of these $\sim 900$~SGRBs,
the $E_p$--$L_p$ relation for SGRBs found by Tsutsui et al.~\cite{Tsutsui:2012dy}
can be used.
Here, $E_p$ and $L_p$ mean the peak energy of the photon and 
the peak luminosity of the SGRB, respectively.
The empirical relation is given by
\begin{equation}
 L_{\rm p} = 10^{52.29 \pm 0.066}~{\rm erg~s^{-1}}
 \left( \frac{E_{\rm p}}{\rm 774.5~keV} \right)^{1.59 \pm 0.11} \,.
\end{equation}
Using the observed flux $f_{\rm p}=L_{\rm p}/(4\pi d_L(z)^2$)
with the luminosity distance $d_L(z)$,
and the peak energy of the photon $E_{\rm p}^{\rm obs}=E_{\rm p}/(1+z)$,
we can determine the redshift $z$.
Yonetoku et al.~\cite{Yonetoku:2014fua}
used the 72 bright BATSE SGRBs for the analysis
since we need many photons to determine $E_p$.
They obtained the minimum event rate of SGRBs
as ${\rm 1.15^{+0.57}_{-0.71}\times 10^{-7}~events~Mpc^{-3}~yr^{-1}}$,
with the observational input that the mean jet opening angle of SGRBs is $6^\circ$.
This corresponds to the minimum event rate of ${\rm 3.9^{+1.9}_{-2.4}~events~yr^{-1}}$
if SGRBs are NS--NS mergers, 
while the minimum event rate is ${\rm 152^{+75}_{-94}~events~yr^{-1}}$
if SGRBs are the NS--BH (of mass 10$M_\odot$) mergers.
In this analysis, because they do not use NS--NS binary data,
even if only $10\%$ of SGRBs
are NS--BH binaries, the minimum event rate becomes
${\rm \sim 20~events~yr^{-1}}$
(see also Refs.~\cite{Petrillo:2012ij,Nissanke:2012dj,Siellez:2013hia,Regimbau:2014nxa}
for the related works).

The third method is the theoretical population synthesis method. 
A recent paper by Dominik et al.~\cite{Dominik:2014yma} includes
the chemical evolution effect of galaxies with metalicity from $1\%$ to $0.01\%$,
that is, from Pop I to Pop II stars (see Ref.~\cite{Belczynski:2014iua}
for a formation of massive stellar BH-BH binaries).
The resulting event rates depend on various parameters, assumptions on the binary interaction,
detectors and the number of detectors, to yield $0.3$--$7$~${\rm events~yr^{-1}}$
for NS--NS mergers and $0.007$--$9.2$~${\rm events~yr^{-1}}$ for NS--BH mergers
(see also Ref.~\cite{Abadie:2010cf} for predictions for the event rates
of compact binary coalescences).

In the case of BH--BH binaries, there is no definite candidate
so that the theoretical population synthesis is the only method to predict the rate.
In this regard, Pop III stars are important.
Pop III stars are the first stars in our universe without the heavy metal,
that is, their envelope consists of H and He only. While Pop I stars are similar to
our sun with $\sim 2\%$ heavy elements in mass with atomic number larger than carbon.
Pop II stars are low metal stars with $\sim 0.01\%$ heavy elements in mass.
Observationally Pop I and Pop II stars suffer mass loss
due to the absorption of photons at the metal lines.
This means that because Pop III stars do not lose their mass,
massive BHs are more likely to be formed.

Recently Kinugawa et al.~\cite{Kinugawa:2014zha,Kinugawa:2015}
showed by using the population synthesis code
that the typical chirp mass of the Pop III binary BHs is $\sim 30 M_\odot$
with the total mass of $\sim 60 M_\odot$.
This means that comparable mass ratio binary BHs are typical.
Here, we note three important facts.
The first one is that Pop III stars were 
formed at $z \sim 10$, while a sizable fraction of BH--BH binaries merge today
since the merger time is proportional to the fourth power
of the initial orbital separation.
The second one is that the typical mass of Pop III stars had been considered
to be $\sim 1000 M_\odot$, while Hosokawa et al.~\cite{Hosokawa:2011qa}
showed that the UV photons from the central star 
evaporate the accretion disk
so that the mass is around $40M_\odot$.
The third one is that Pop III star of mass $\sim 30M_\odot$ ends its life
not as a red giant but as a blue giant,
according to the evolution calculations of Pop III stars
by Marigo et al.~\cite{Marigo:2001pm}.
This means that since the mass loss due to the binary interaction is small,
the formation of $30M_\odot$--$30M_\odot$ binary BHs is expected.
Kinugawa et al.~\cite{Kinugawa:2014zha,Kinugawa:2015} performed
the population synthesis of $10^6$ binary Pop III stars for 14 models
and obtained 
$14.6$--$599.3$~${\rm events~yr^{-1}~(SFR_p/(10^{-2.5}~M_\odot~yr^{-1}~Mpc^{-3}))}
\cdot (\rm [f_b/(1+f_b)]/0.33)$ for aLIGO~\cite{TheLIGOScientific:2014jea},
AdV~\cite{TheVirgo:2014hva} and KAGRA~\cite{Somiya:2011np,Aso:2013eba}
where ${\rm SFR_p}$ and ${\rm f_b}$ are the peak value of the Pop III star
formation rate and the fraction of binaries, respectively.
We should emphasize here that the factor $\sim 40$ difference
among various models exists like
in Refs.~\cite{Kim:2002uw,Kalogera:2003tn,Kim:2013tca} for NS--NS merger rates.
In the cases of Pop I and Pop II, Dominik et al.~\cite{Dominik:2014yma}
showed that the typical chirp mass of BH--BH binary is smaller than
that in the Pop III case
and the event rate ranges from $0.6$ to $1338$~${\rm events~yr^{-1}}$.

As shown by Kanda et al.~\cite{LCGT:2011aa},
the chirp signal and the quasinormal mode (QNM) frequency
of $\sim 30M_\odot$--$30M_\odot$ binary BHs
are in the best sensitivity band of aLIGO, AdV and KAGRA.
Therefore, there is a good chance to observe the QNM
which will exhibit the strong gravity space-time near the event horizon
to confirm Einstein's general relativity.
To confirm the expected signal of the QNM,
the threshold signal-to-noise ratio (SNR)~$=8$
which is usually used for the detection, is not enough.
For this confirmation, Nakano, Tanaka and Nakamura~\cite{Nakano:2015uja}
have shown that SNR $\sim 35$ events of $30 M_\odot$--$30M_\odot$
Pop III BH--BH merger are the appropriate target.
Then, for the Pop III case, the event rate for SNR $\gtrsim 35$ becomes
$0.17$--$7.2$~${\rm events~yr^{-1}~(SFR_p/(10^{-2.5}~M_\odot~yr^{-1}~Mpc^{-3}))}
\cdot (\rm [f_b/(1+f_b)]/0.33)$ (see Ref.~\cite{Belczynski:2015tba}
for a comparison of evolutionary predictions with initial and forthcoming
LIGO/Virgo upper limits, and related works~\cite{Dominik:2012kk,Dominik:2013tma}). 
It is noted that there are various proposals for the formation of heavy BH
binaries~\cite{Amaro-Seoane:2015umi,Mandel:2015qlu,Marchant:2016wow}.

What we can say definitely 
when the QNM is confirmed to exist at the expected frequency,
is the main theme of this paper.
Since QNMs are obtained under the ingoing and outgoing wave conditions
at the event horizon and the spatial infinity, respectively,
the most optimistic statement is that all the space-time of a Kerr BH is tested.
However, even if the event horizon might be absent, fuzzy or blocked
by the firewall (see, e.g.,
Refs.~\cite{Mazur:2001fv,Mathur:2005zp,Braunstein:2009my,Almheiri:2012rt}),
we will observe something similar to the QNMs predicted by general relativity
(see Refs.~\cite{Damour:2007ap,Barausse:2014tra}).
In this paper, we would like to ask what we can say more robustly.
 
This paper is organized as follows. In Sec.~\ref{sec:RWZ},
we compare the Regge-Wheeler and Zerilli potentials~\cite{Regge:1957td,Zerilli:wd}
and the WKB analysis of the QNM for the Schwarzschild BH.
In Sec.~\ref{sec:SN}, we use the Sasaki-Nakamura
equation~\cite{Sasaki:1981kj,Sasaki:1981sx,Nakamura:1981kk} for the Kerr BH,
while in Sec.~\ref{sec:CD}, we use the (Chandrasekhar and)
Detweiler equation~\cite{Detweiler:1977gy}.
Section~\ref{sec:dis} is devoted to discussions. 
In the analysis of QNMs in this paper, we focus only on the ($\ell=2,\,m=2$) mode
since BH binary merger simulations indicate that
the ($\ell=2,\,m=2$) QNM is dominant (see e.g., Ref.~\cite{London:2014cma}).
We use the geometric unit system, where $G=c=1$ in this paper.

%%%%%%%%%%%%%%%%%%%%%%%%%%%%%%%%%%%%%%%%
\section{Regge-Wheeler and Zerilli Equation}\label{sec:RWZ}
%%%%%%%%%%%%%%%%%%%%%%%%%%%%%%%%%%%%%%%%

For the Schwarzschild BH case with mass $M$, the metric is given by
\bea
 ds^2 = -\left(1-\frac{2M}{r}\right) dt^2
 + \left(1-\frac{2M}{r}\right)^{-1} dr^2
 + r^2 (d\theta^2 + \sin^2 \theta d\phi^2) \,.
\eea
We may use the Regge-Wheeler
and Zerilli equations~\cite{Regge:1957td,Zerilli:wd}
to obtain the gauge invariant perturbation.
The Regge-Wheeler (odd parity) / Zerilli (even parity) function $\psi^{\rm RW/Z}$
satisfies the Regge-Wheeler or Zerilli equation,
\begin{eqnarray}
 \frac{d^2 \psi^{\rm RW/Z}}{{d r^*}^2} 
 + \left(\omega^2 - V_{\rm RW/Z} \right)
 \psi^{\rm RW/Z} = 0 \,,
 \label{eq:RWZ}
\end{eqnarray}
where $r^*=r+2M \log (r/2M-1)$ and the potentials are given as
\begin{eqnarray}
 V_{{\rm RW}}
 &=&\left(1-\frac{2M}{r}\right)
 \left[ \frac{\ell(\ell+1)}{r^2}-\frac{6M}{r^3} \right]
 \,, \cr
 V_{{\rm Z}}
 &=& 
 \frac {r-2M}{{r}^{4} ( r{\ell}^{2}+r\ell-2\,r+6\,M ) ^{2}}
 \nonumber \\ && \times 
 \left[ \ell ( \ell+1 )  ( \ell+2 ) ^{2} ( \ell-1 ) ^{2}{r}^{3}
 +6\,M ( \ell+2 ) ^{2} ( \ell-1 ) ^{2}{r}^{2}
 +36\,{M}^{2} ( \ell+2 )  ( \ell-1 ) r+72\,{M}^{3} \right] \,,
\end{eqnarray}
where $\ell$ is the index of the spherical harmonics $Y_{\ell m}(\theta, \phi)$ .
The potential for the even parity is slightly complicated, but
according to Ref.~\cite{Chandrasekhar:1985kt},
both $V_{\rm Z}$ and $V_{\rm RW}$ are written in a unified manner as
\begin{eqnarray}
 V_{\rm Z/RW}(r) &=&
 \pm \beta \frac{d y}{dr^*} + \beta^2 y^2 + \kappa y \,;
 \cr
 \beta &=& 6M \,, \quad \kappa = (\ell-1)\ell(\ell+1)(\ell+2) \,,
 \quad y = \frac{r-2M}{r^2( r{\ell}^{2}+r\ell-2\,r+6\,M )} \,,
\end{eqnarray}
where the upper and lower signs are for the even and odd parities, respectively.
Then, the Regge-Wheeler and Zerilli equations become
\begin{eqnarray}
 \frac{d^2 \psi^{\rm Z/RW}}{{d r^*}^2}
 = \left( - \omega^2 \pm \beta \frac{d y}{dr^*} + \beta^2 y^2 + \kappa y \right)
 \psi^{\rm Z/RW} \,.
\end{eqnarray}
For the even parity, we can have the same differential equation
as the odd parity, i.e., 
the modified function $\psi^{\rm even,RW}$ satisfies 
\begin{eqnarray}
 \frac{d^2 \psi^{\rm even,RW}}{{d r^*}^2}
 = \left( - \omega^2 - \beta \frac{d y}{dr^*} + \beta^2 y^2 + \kappa y \right)
 \psi^{\rm even,RW} \,.
\end{eqnarray}
To do so, we consider the transformation of the wave function,
\begin{eqnarray}
 \psi^{\rm Z} &=&
 \frac{1}{C} \left[
 \left( \frac{\kappa}{4} + \frac{\beta^2}{2} y \right)
 \psi^{\rm even,RW}
 + \frac{\beta}{2} \frac{d \psi^{\rm even,RW}}{d r^*} \right]
 \,,
 \cr
 \psi^{\rm even,RW} &=&
 \left( \frac{\kappa}{4} + \frac{\beta^2}{2} y \right)
 \psi^{\rm Z}
 - \frac{\beta}{2} \frac{d \psi^{\rm Z}}{d r^*} \,,
\end{eqnarray}
where
\begin{eqnarray}
 C &=& \frac{\beta^2 \omega^2}{4} + \frac{\kappa^2}{16} \,.
\end{eqnarray}
This is called the Chandrasekhar transformation.
Therefore, the same QNMs are obtained
in the Regge-Wheeler and Zerilli equations.

To study BH QNMs, the WKB approximation has been used
frequently~\cite{Mashhoon:1985cya}.
While the full analysis of the QNMs gives the same frequencies
for the odd and even parity perturbations, we have different results
in the leading order WKB analysis~\cite{Schutz:1985zz}
(see Sec.~\ref{sec:SN} for a brief summary).
In the WKB approximation, the QNM frequency is determined by
the information around the peak of the potential.
The peak of $V_{\rm RW}$ is obtained explicitly as
\bea
 r^{\rm RW}_0 = 
 \frac{1}{2}\,{\frac {M \left( 3\,{\ell}^{2}+3\,\ell+9+
 \sqrt {9\,{\ell}^{4}+18\,{\ell}^{3}-33\,{\ell}^{2}-42\,\ell+81} \right) }
 {\ell \left( \ell+1 \right) }} \,.
\eea
On the other hand, we calculate the peak location of $V_{\rm Z}$
in the large $\ell$ expansion because of the complicated potential,
and derive
\bea
 r^{\rm Z}_0 = 
 3\,M \left( 1+\frac{1}{3}\,\frac{1}{\ell^{2}} -\frac{1}{3}\,\frac{1}{\ell^{3}} 
 -\frac{1}{9}\,\frac{1}{\ell^{4}} + O(\ell^{-5}) \right) \,.
\eea
This should be compared with $r^{\rm RW}_0$ in the large $\ell$ expansion,
\bea
 r^{\rm RW}_0 = 
 3\,M \left( 1+\frac{1}{3}\,\frac{1}{\ell^{2}} -\frac{1}{3}\,\frac{1}{\ell^{3}} 
 +\frac{11}{9}\,\frac{1}{\ell^{4}} + O(\ell^{-5}) \right) \,,
\eea
which shows that
the difference between $r^{\rm RW}_0$ and $r^{\rm Z}_0$ is $O(\ell^{-4})$.
For example, this difference for the $\ell=2$ mode is numerically shown in
\bea
 r^{\rm RW}_0 \approx 3.28077M \,, \quad r^{\rm Z}_0 \approx 3.09879 M \,.
\eea
Therefore, although the peak location slightly depends on which potential
we adopt, this can be a good estimator to discuss where the QNMs are emitted.

%%%%%%%%%%%%%%%%%%%%%%%%%%%%%%%%%%%%%%%%
\section{Sasaki-Nakamura Equation}\label{sec:SN}
%%%%%%%%%%%%%%%%%%%%%%%%%%%%%%%%%%%%%%%%

For the Kerr BH case, the metric is given by
\bea
 ds^2 = - \left( 1 - \frac{2 M r}{\Sigma} \right) dt^2 
 - \frac{4 M a r \sin^2 \theta}{\Sigma} dt d\phi 
 + \frac{\Sigma}{\Delta} dr^2 + \Sigma d\theta^2 
 + \left( r^2 + a^2 + \frac{2 M a^2 r}{\Sigma} \sin^2 \theta \right)
 \sin^2 \theta d\phi^2 \,,
\eea
where 
$M$ and $a$ are the mass and the spin parameter, respectively, 
$\Sigma = r^2 + a^2 \cos^2 \theta$ and $\Delta = r^2 - 2 M r + a^2$.
Teukolsky~\cite{Teukolsky:1973ha} showed that the gravitational perturbation
is separable and the radial equation is given by
\begin{eqnarray}
 \Delta^2\frac{d}{dr}\frac{1}{\Delta}\frac{dR}{dr}-VR = -T \,,
\end{eqnarray}
where $T$ is the source and
\begin{eqnarray}
 V = -\frac{K^2}{\Delta}-\frac{2iK\Delta'}{\Delta}+4iK'+\lambda \,,
\end{eqnarray}
with
\begin{eqnarray}
 K = (r^2+a^2)\,\omega - am \,,
 \label{eq:K}
\end{eqnarray}
and $m$ and $\lambda$ come from $e^{im\phi}$
and the spin-weighted spheroidal function $Z^{a\omega}_{lm}(\theta)$,
respectively.
Here, a prime means the derivative with respect to $r$.
The source term $T$ diverges
as $\propto r^{7/2}$ for a particle falling into the Kerr BH,
and the potential $V$ takes the long range nature.
Therefore, Sasaki and Nakamura~\cite{Sasaki:1981kj,Sasaki:1981sx,Nakamura:1981kk}
proposed a new equation with the convergent source term and
the short range potential as the followings.

Specifying two functions $\alpha(r)$ and $\beta(r)$,
we can define various variables as
\begin{eqnarray}
 X &=& \frac{\sqrt{r^2+a^2}}{\Delta}
 \left(\alpha R+\frac{\beta}{\Delta}R'\right) \,,
 \label{eq:XR} \\
 \gamma &=& \alpha \left(\alpha+\frac{\beta'}{\Delta}\right)
 -\frac{\beta}{\Delta}\left(\alpha'+\frac{\beta}{\Delta^2}V\right) \,,
 \label{eq:gamma} \\
 F &=& \frac{\Delta}{r^2+a^2}\frac{\gamma'}{\gamma} \,, \cr
 U_0 &=& V+\frac{\Delta^2}{\beta}
 \left[ \left(2\alpha+\frac{\beta'}{\Delta}\right)'
 -\frac{\gamma'}{\gamma}\left(\alpha+\frac{\beta'}{\Delta}\right)\right] \,,
 \label{eq:U0} \\
 G &=& -\frac{\Delta'}{r^2+a^2}+\frac{r\Delta}{(r^2+a^2)^2} \,,
 \label{eq:G} \\
 U &=& \frac{\Delta U_0}{(r^2+a^2)^2}+G^2+\frac{dG}{dr^*}
 -\frac{\Delta G}{r^2+a^2}\frac{\gamma'}{\gamma} \,.
 \label{eq:U}
\end{eqnarray}
Then, we have the wave equation as 
\begin{equation}
 \frac{d^2X}{dr^{*2}}-F\frac{dX}{dr^*}-UX=0 \,,
\end{equation}
where
\bea
 \frac{dr^*}{dr} &=& \frac{r^2+a^2}{\Delta} \,.
\eea

We adopt $\alpha$ and $\beta$ defined by
\begin{eqnarray}
 \alpha &=& A-\frac{iK}{\Delta} B \,, \cr
 \beta &=& \Delta B \,,
\end{eqnarray}
where
\begin{eqnarray}
 A &=& 3iK'+\lambda+6 \frac{\Delta}{r^{2}} \,, \cr
 B &=& -2iK+\Delta'-4\frac{\Delta}{r} \,.
\end{eqnarray}
Then, defining a new variable $Y$ by $X=\sqrt{\gamma}~Y$, we have
\begin{eqnarray}
 \frac{d^2Y}{dr^{*2}}+\left( \omega^2-V_{\rm SN} \right) Y =0 \,,
\end{eqnarray}
where
\begin{eqnarray}
 V_{\rm SN}=\omega^2+U
 -\left[ \frac{1}{2}\frac{d}{dr^*}
 \left(\frac{1}{\gamma}\frac{d\gamma}{dr^*}\right)
 -\frac{1}{4\gamma^2}\left(\frac{d\gamma}{dr^*}\right)^2\right] \,.
\end{eqnarray}
A remarkable feature of $V_{\rm SN}$
is that the potential becomes the Regge-Wheeler potential for $a=0$.

Schutz and Will~\cite{Schutz:1985zz}
derived the QNM for the Regge-Wheeler potential $V_{\rm RW}$
by using the WKB method. 
The essence of their method is to approximate $V_{\rm RW}$
by the expansion near its peak at $r_0 \approx 3.28M$ as 
\begin{eqnarray}
 V_{\rm RW}(r^*) = V_{\rm RW}(r^*_0)
 +\frac{1}{2} \left. \frac{d^2V_{\rm RW}}{dr^{*2}}
 \right|_{r^*=r^*_0}(r^*-r^*_0)^2 \,,
\end{eqnarray}
where $r^*_0=r_0+2M\log(r_0/2M-1)$.
Then the fundamental ($n=0$) QNM frequency is expressed as
\begin{equation}
 (\omega_r+i\omega_i)^2
 = V_{\rm RW}(r^*_0)-i\sqrt{-\frac{1}{2}
 \left. \frac{d^2V_{\rm RW}}{dr^{*2}} \right|_{r^*=r^*_0}} \,.
\end{equation}
They have found that for the $n=0$ QNM frequency
of the $\ell=2$ mode, 
the errors of the real ($\omega_r = {\rm Re}(\omega)$)
and the imaginary ($\omega_i = {\rm Im}(\omega)$) parts
are $7\%$ and $0.7\%$, respectively, compared with the numerical results
of Chandrasekhar and Detweiler~\cite{Chandrasekhar:1975zza}.

Since one needs to impose the ingoing and outgoing boundary conditions
at the event horizon and the spatial infinity, respectively, 
to derive QNM frequencies, 
one might say that the existence of the event horizon is confirmed
if the QNM of a Schwarzschild BH is detected.
However, the WKB method by Schutz and Will~\cite{Schutz:1985zz} suggests
that what we can conservatively claim is that the space-time of a Schwarzschild BH
around $r \approx 3.28M$ is tested
since both real and imaginary parts of the QNM frequency
are determined by the value of $V_{\rm RW}$
and its curvature at $r \approx 3.28M$.

What we want to do is to establish a similar statement for the Kerr BH case.
Since numerical relativity simulations for binary
BHs~\cite{Pretorius:2005gq, Campanelli:2005dd, Baker:2005vv}
suggest that the final Kerr parameter $a/M$
after the merger of equal-mass, nonspinning BHs is about $0.7$
(see Ref.~\cite{Healy:2014yta} for the latest remnant spin formula),
instead of the Regge-Wheeler potential $V_{\rm RW}$
we need to discuss the Kerr case with the potential $V_{\rm SN}$.

An important difference of the Kerr case
is that the potential becomes complex even if one uses the real potential
for the real frequency $\omega$ such as Detweiler~\cite{Detweiler:1977gy}.
Since the separation constant $\lambda$
of the radial and angular Teukolsky equations depends on
$\omega$, which is complex in the QNM calculation,
the above potential becomes complex.
Therefore, there is no advantage to use the real potential 
in our approach, and the complex nature of $V_{\rm SN}$ for $a\neq 0$
is not an essential drawback.
Our purpose is not to determine the QNM frequencies correctly
by the WKB method since there is a good numerical method to determine them
by Leaver~\cite{Leaver:1985ax}.
Our interest is in what we can claim conservatively
when the QNM of a Kerr BH is detected
by the second generation gravitational wave detectors, aLIGO, AdV and KAGRA.
That is, our goal is to find a similar physical picture
as Schutz and Will~\cite{Schutz:1985zz} did for the Schwarzschild case,
i.e., the BH space-time near $r=3.28M$ can be confirmed by the QNMs
of a BH with $a/M=0$.

\begin{figure}[!ht]
\begin{center}
 \includegraphics[width=0.48\textwidth,clip=true]{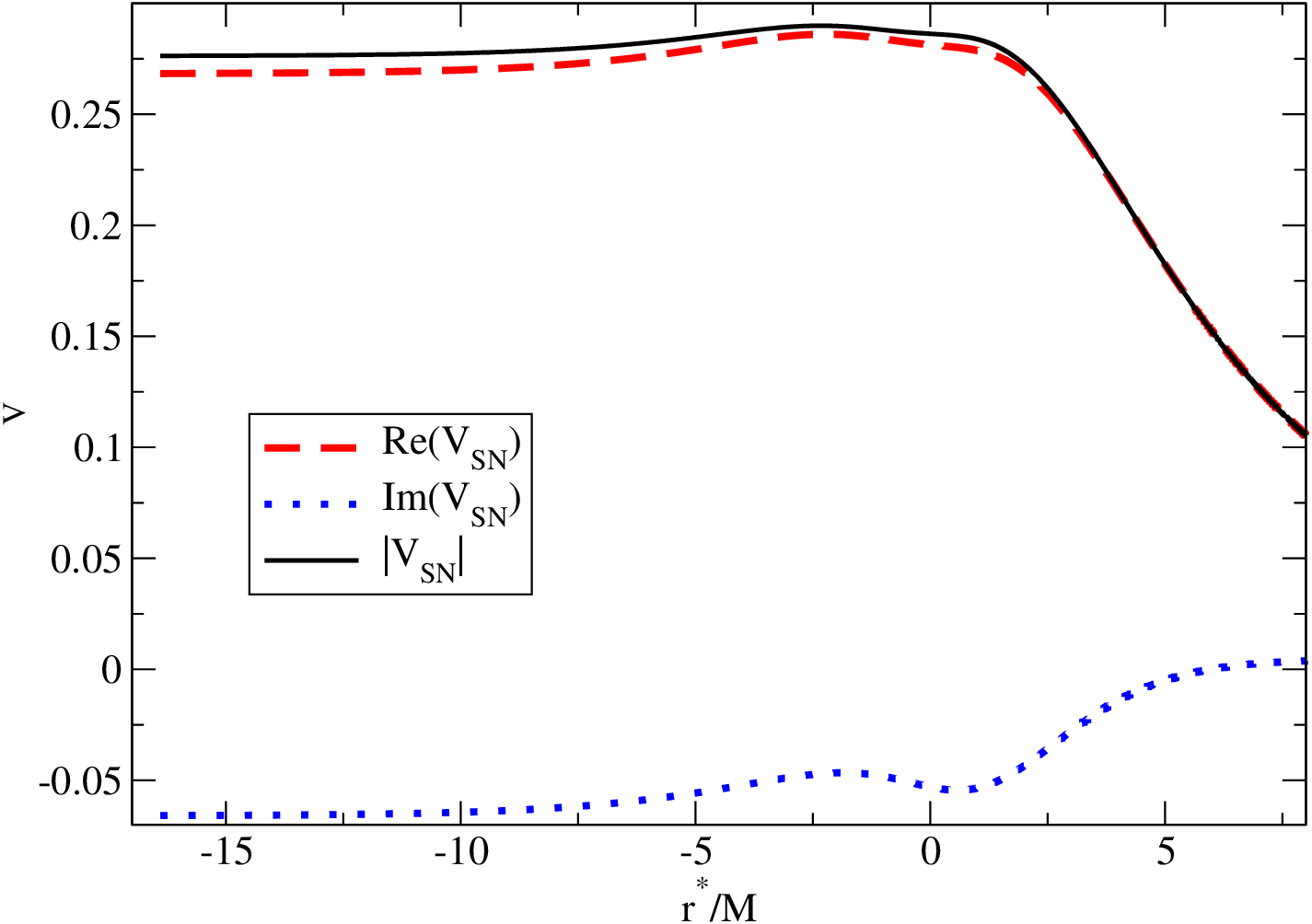}
 \includegraphics[width=0.48\textwidth,clip=true]{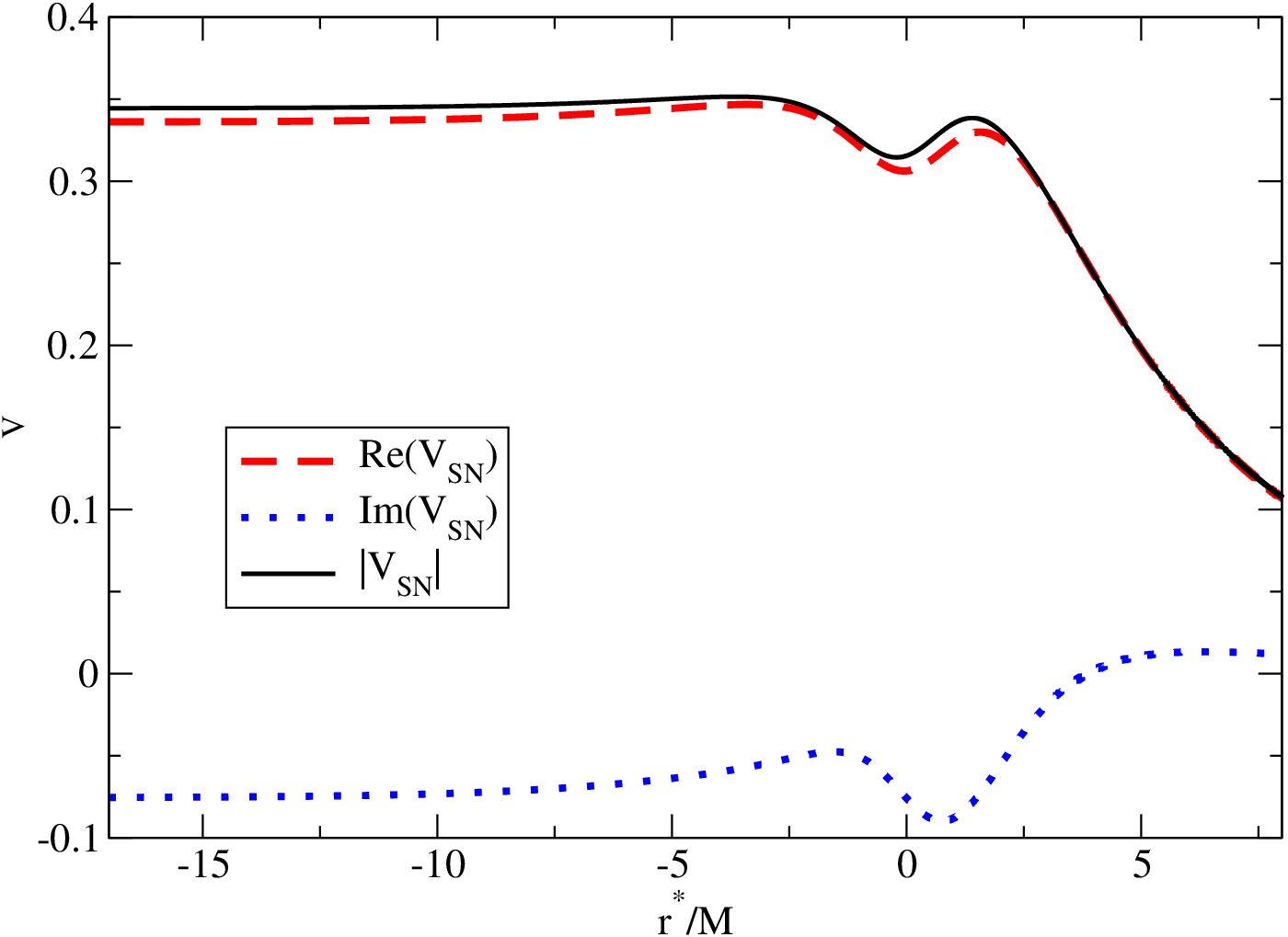}
\end{center}
 \caption{${\rm Re}(V_{\rm SN})$, ${\rm Im}(V_{\rm SN})$ and $|V_{\rm SN}|$
 with ($\ell=2,\,m=2$)
 for $a/M=0.7$ (left) and $0.8$ (right) as a function of $r^*/M$
 where we set $M=1$.}
 \label{fig:VSN0708}
\end{figure}

We therefore substitute the numerical value of the QNM frequency
for the Kerr BH case by Ref.~\cite{Berti:2005ys}
(see useful Berti's ``Ringdown'' website~\cite{BertiQNM})
as well as the $a\omega$ dependence of $\lambda$ up to the sixth order
by Ref.~\cite{Press:1973zz} into the potential $V_{\rm SN}$.
It is noted that the error between the fitting function of lambda and
the exact numerical value is less than $0.015\%$ for the QNM frequency.
The left panel of Fig.~\ref{fig:VSN0708} shows the values of the potential,
${\rm Re}(V_{\rm SN})$, ${\rm Im}(V_{\rm SN})$ and $|V_{\rm SN}|$
with $(\ell=2,\,m=2$) as a function of $r^*/M$ in the case of $a/M=0.7$.
We see that $|V_{\rm SN}|$
is close to ${\rm Re}(V_{\rm SN})$ and contribution of ${\rm Im}(V_{\rm SN})$ is small
so that we can identify the critical radius $r^* \sim -2M$, i.e., $r \sim 2M$
with the peak of  $|V_{\rm SN}|$.
For the $a/M=0.8$ case with $(\ell=2,\,m=2$), 
it is still reasonable to use the peak
as shown in the right panel of Fig.~\ref{fig:VSN0708}.
Since we cannot apply the standard WKB approximation
for double peaks shown in the right panel of Fig.~\ref{fig:VSN0708},
the estimation for $a/M=0.8$ in the current analysis
is not mathematically appropriate.
For smaller values of $a/M$, the contribution of
${\rm Im}(V_{\rm SN})$ is much smaller than the $a/M=0.7$ case. 

\begin{figure}[!ht]
\begin{center}
 \includegraphics[width=0.48\textwidth,clip=true]{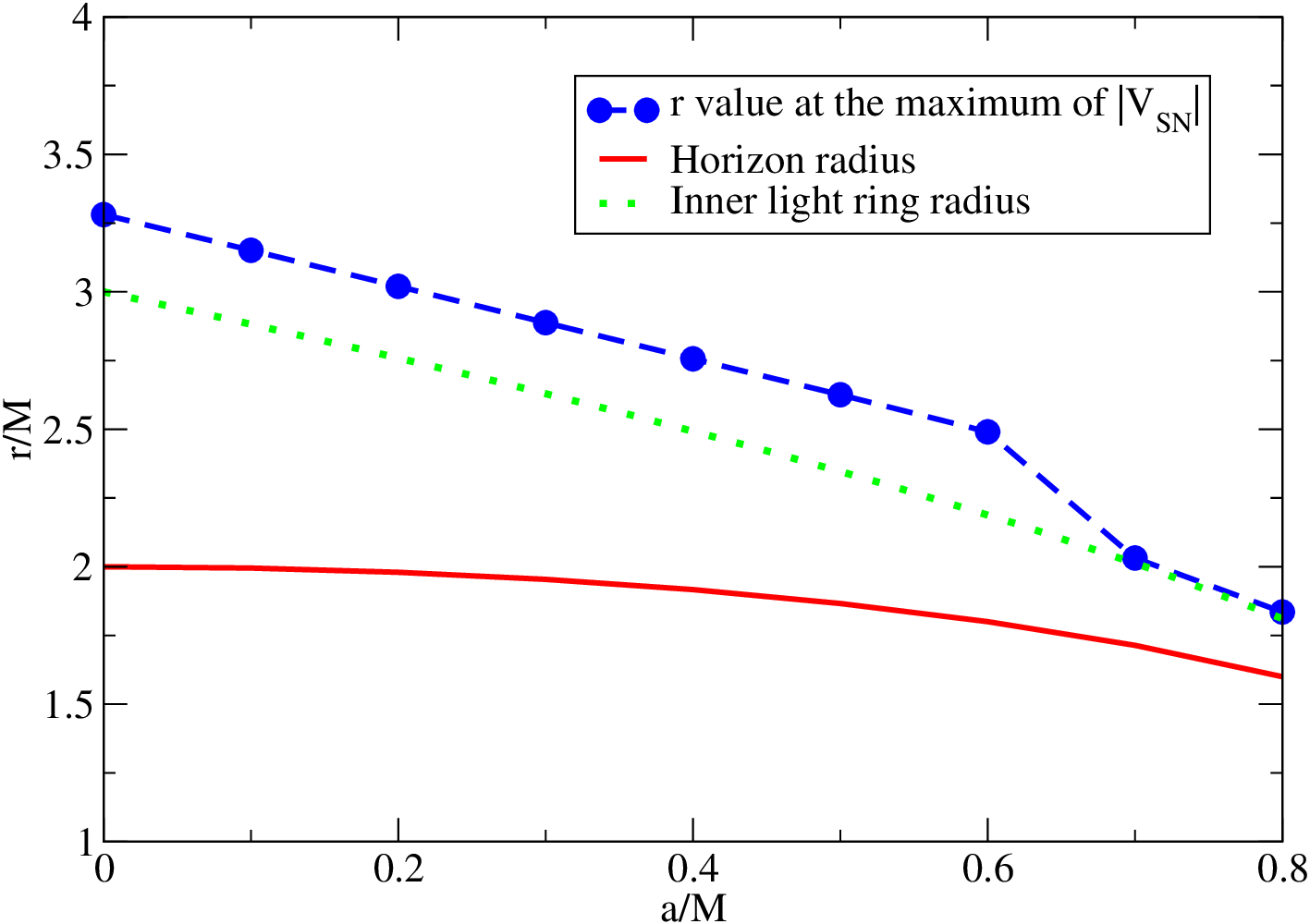}
\end{center}
 \caption{The location of the maximum of the absolute value of the potential
 $|V_{\rm SN}|$ with ($\ell=2,\,m=2$), 
 the event horizon $r_{+}=M + \sqrt{M^2-a^2}$,
 and the inner light ring radius
 $r_{\rm lr}=2M(1+\cos((2/3)\cos^{-1}(-a/M)))$~\cite{Bardeen:1972fi}
 evaluated for various spin parameters $a$.}
 \label{fig:location}
\end{figure}

\begin{figure}[!ht]
\begin{center}
 \includegraphics[width=0.48\textwidth,clip=true]{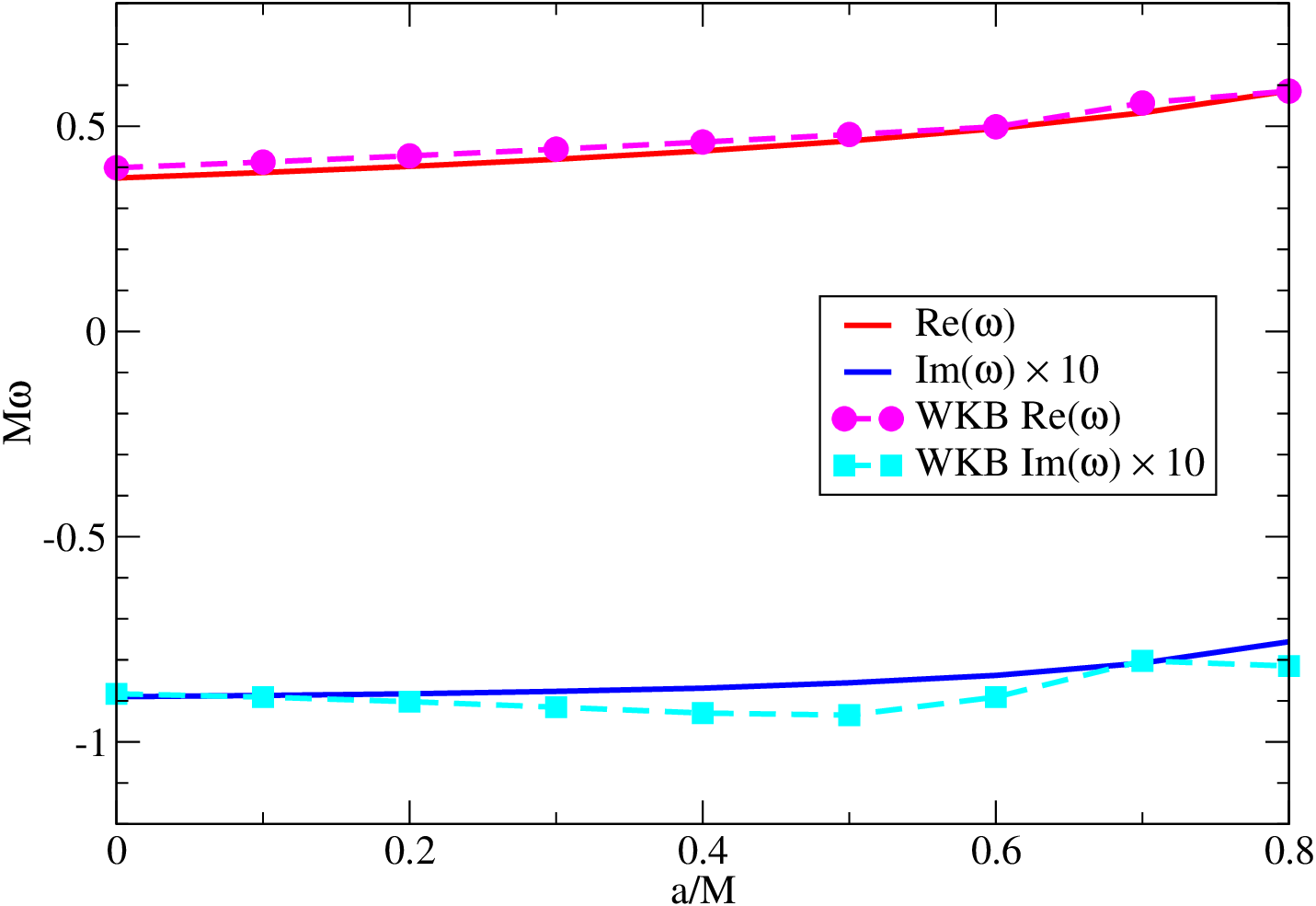}
\end{center}
 \caption{The real and imaginary parts of the fundamental ($n=0$)
 QNM frequencies with ($\ell=2,\,m=2$) evaluated for various spin parameters $a$.
 The exact frequencies ${\rm Re}(\omega)$ and ${\rm Im}(\omega)$
 are from Ref.~\cite{Berti:2005ys}.}
 \label{fig:frequencies}
\end{figure}

\begin{figure}[!ht]
\begin{center}
 \includegraphics[width=0.48\textwidth,clip=true]{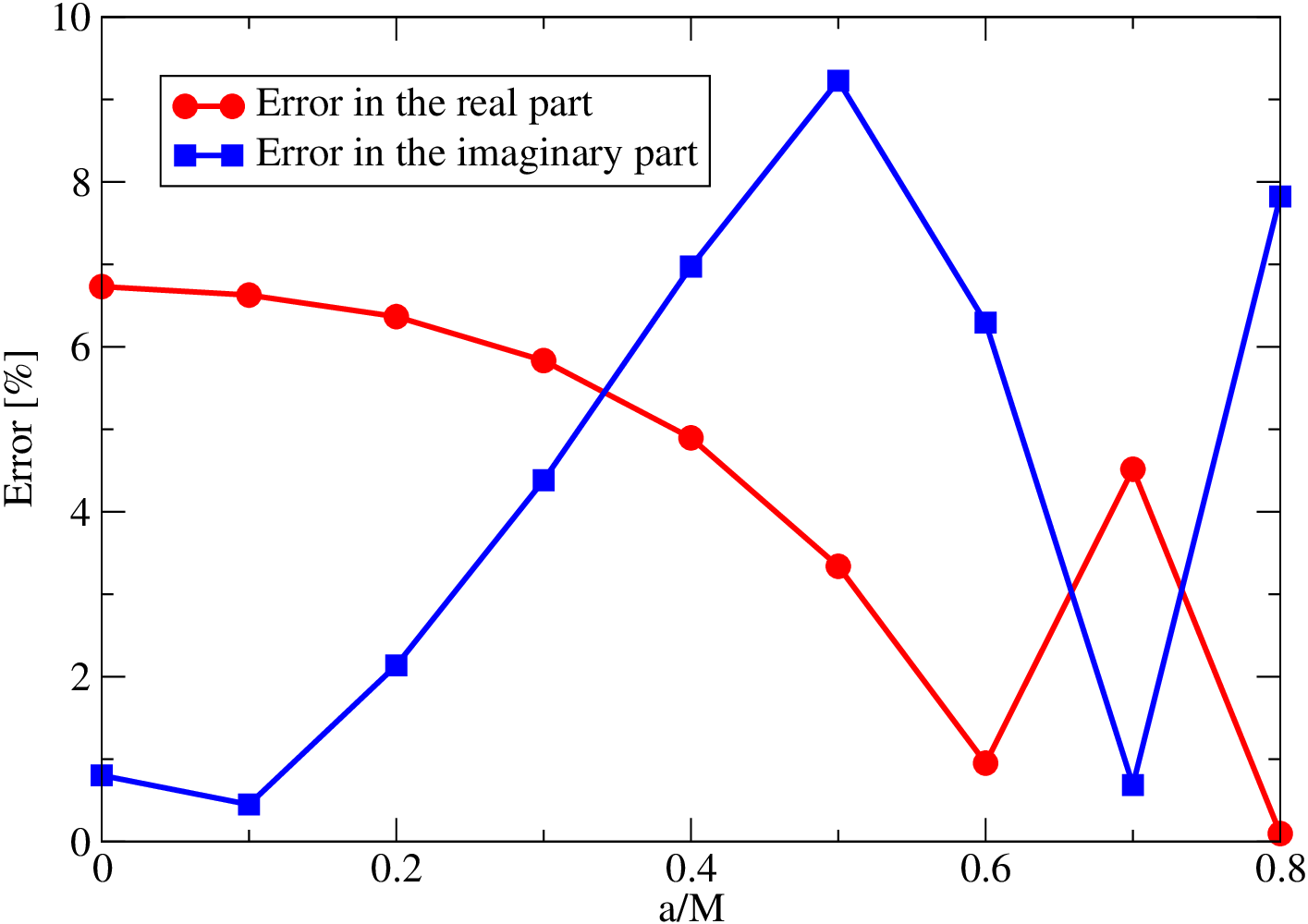}
\end{center}
 \caption{Absolute value of relative errors for the QNM frequencies with ($\ell=2,\,m=2$)
  between the exact value and that of the WKB approximation in Fig.~\ref{fig:frequencies}.}
 \label{fig:errors}
\end{figure}

Therefore, it would not be a bad approximation to expand $V_{\rm SN}$
near the peak of $|V_{\rm SN}|$ (or ${\rm Re}(V_{\rm SN})$)
as Schutz and Will~\cite{Schutz:1985zz} 
did for the Schwarzschild case as
\begin{equation}
 V_{\rm SN}(r^*) = V_{\rm SN}(r^*_0)
 +\frac{1}{2} \left. \frac{d^2V_{\rm SN}}{dr^{*2}}\right|_{r^*=r^*_0}
 (r^*-r^*_0)^2 \,.
 \label{eq:2ndVsn}
\end{equation}
For definiteness, we choose to evaluate the 
above expression at the peak of $|V_{\rm SN}|$, and 
we denote that real-valued radius as $r^*_0$. 
Then, we can estimate the $n=0$ QNM frequency by
\begin{equation}
 (\omega_r+i\omega_i)^2
 = V_{\rm SN}(r^*_0) -i\sqrt{-\frac{1}{2}
 \left. \frac{d^2V_{\rm SN}}{dr^{*2}}\right|_{r^*=r^*_0}} \,.
\label{eq:omega}
\end{equation}
This approximation will be worse when ${\rm Im}(V_{\rm SN})$ becomes large.
The location of the maximum of the absolute value of the potential $|V_{\rm SN}|$,
and the real and imaginary parts of the $n=0$ QNM frequencies
with ($\ell=2,\,m=2$)
are shown in Figs.~\ref{fig:location} and \ref{fig:frequencies}, respectively.
In Fig.~\ref{fig:errors}, we present the errors in the frequencies.
We can say that our approximate method reproduces the QNM frequencies
within $10\%$ accuracy which supports our approximation.
It is noted that fine structures in the imaginary part
of Figs.~\ref{fig:location} and \ref{fig:frequencies}
are due to the complicated behavior of the potential with respect to $a/M$.
A local maximum arises in ${\rm Im}(V_{\rm SN})$ for $a/M \gtrsim 0.6$,
and ${\rm Re}(V_{\rm SN})$ has two peaks for $a/M \gtrsim 0.8$.

In the above analysis, 
assuming the smallness of ${\rm Im}(V_{\rm SN})$,
we have derived $r^*_0$ by finding the peak of $|V_{\rm SN}|$.
But to use Eq.~\eqref{eq:2ndVsn} exactly,
it is necessary to find $r^*_0$ for $dV_{\rm SN}/dr^*=0$
in the complex plane. Then, we may mention the effective peak radius
by the real part of $r^*_0$
because the imaginary part of $r^*_0$ is small.
This also supports the argument to use the real $r^*_0$ 
derived from the peak of $|V_{\rm SN}|$.
Some brief analysis is summarized in Appendix~\ref{app:ano},
and we find a good agreement between the peak location of $|V_{\rm SN}|$
and the real part of $r^*_0$ of $dV_{\rm SN}/dr^*=0$.

As a result, we may say that if the QNM with $a/M\sim 0.7$ is detected
with an accuracy of $10\%$, the BH space-time around $r\sim 2M$
which is $1.17$ times the event horizon, is confirmed.
Here, we know that we cannot localize gravitational waves within 
a small region less than its wavelength. 
Since the wavelength of the QNM that we are concerned 
is $O(10M)$, the origin of the QNM is also necessarily extended 
to a similar amount. The extension should be measured in the $r^*$ 
coordinate. The non-local nature of gravitational waves 
is reflected by the appearance of the second derivative of the 
potential in the expression of Eq.~\eqref{eq:omega}. The local value 
of the second derivative at the peak can be varied by arranging the 
transformation. However, we would have to pay the expense 
that the potential will tend to take more wavy shape and 
possibly possess multiple extrema. 

From the Pop III star population synthesis, the expected event rate is
$0.17$--$7.2$~${\rm events~yr^{-1}~(SFR_p/(10^{-2.5}~M_\odot~yr^{-1}~Mpc^{-3}))}
\cdot (\rm [f_b/(1+f_b)]/0.33)$ where ${\rm SFR_p}$ and ${\rm f_b}$
are the peak value of the Pop III star formation rate and the fraction
of binaries, respectively~\cite{Kinugawa:2015}.
Since the range of the above rate is not derived from the statistical treatment
and the minimum rate of $0.17$ is obtained from the 
most unlikely model,
there will be a good chance to observe the QNM with SNR $\sim 35$.

%%%%%%%%%%%%%%%%%%%%%%%%%%%%%%%%%%%%%%%%
\section{Chandrasekhar and Detweiler Equation}\label{sec:CD}
%%%%%%%%%%%%%%%%%%%%%%%%%%%%%%%%%%%%%%%%

Chandrasekhar and Detweiler developed various transformations
in the radial Teukolsky equation
(see, e.g, Ref.~\cite{Chandrasekhar:1976zz} and related references therein).
Here, we use the notation in Appendix B of Ref.~\cite{Detweiler:1977gy}.

The two functions $\alpha(r)$ and $\beta(r)$ in Eq.~\eqref{eq:XR} are now
\bea
 \alpha &=& \frac{2\sqrt{2}\rho^4}{|\kappa|\Delta}
 \left[ {\cal R} + T^* \left(\frac{3r\Delta}{\rho^4}-i\sigma\right) \right] \,,
 \cr
 \beta &=& -\frac{2\sqrt{2}\Delta\rho^2T^*}{|\kappa|} \,,
 \label{eq:abCD}
\eea
where $\Delta$ is the same as the one in the previous section while $\sigma=-\omega$,
and
\bea
 \rho^2 &=& r^2+a^2+\frac{am}{\sigma} \,, \cr
 {\cal R} &=& \frac{\Delta^2}{\rho^8} (F + b_2) \,, \cr
 T^* &=& -2i\sigma + \frac{1}{F-b_2} \left(\frac{\Delta}{\rho^2}\frac{dF}{dr} - \kappa_2 \right) \,,\\
 \kappa &=& \left[ \lambda^2(\lambda+2)^2 +144 a^2 \sigma^2(a\sigma+m)^2
 -a^2\sigma^2(40\lambda^2-48\lambda)  -a\sigma m(40\lambda^2+48\lambda) \right]^{1/2}
 -12i\sigma M \,.
\eea
There is a typo~\footnote{
Although we do not use Eq.~(B8) of Ref.~\cite{Detweiler:1977gy},
there should be a typo, and it reads as
\bea
 a_2 = \frac{1}{\Delta^2}
 \left[ \frac{24\sigma r K^2}{\Delta} - \frac{4\lambda (r-M)K}{\Delta} -4\sigma r\lambda
 -12\sigma M \right] \,.
\eea
Here, the definition of $K$ in the above equation
has the inverse signature of Eq.~\eqref{eq:K} due to $\sigma=-\omega$.}
in Eq.~(B19) of Ref.~\cite{Detweiler:1977gy}, 
i.e., $-i\sigma$ in $T^*$ should be $-2i\sigma$ as the above equation.
We have confirmed that 
$\alpha(r)$ and $\beta(r)$ in Eqs.~\eqref{eq:abCD}
give a constant $\gamma$ in Eq.~\eqref{eq:gamma},
i.e., a constant ${\cal K}$ in Eq.~(11) of Ref.~\cite{Detweiler:1977gy}.
In the above equations, we have
\bea
 F &=& \frac{1}{\Delta} \left[ \lambda \rho^4 + 3\rho^2 (r^2-a^2) -3r^2\Delta \right] \,,
 \cr
 b_2 &=& \pm 3 \left( a^2 + \frac{am}{\sigma} \right) \,, \label{eq:pmb2} \\
 \kappa_2 &=& \pm \left\{ 36M^2 - 2\lambda \left[ (a^2+am/\sigma)(5\lambda+6)-12a^2 \right]
 +2b_2\lambda (\lambda+2) \right\}^{1/2} \,. \label{eq:pmkappa2}
\label{eq:kappa2}
\eea

The differential equation for $X$ becomes
\bea
 \frac{d^2X}{dr^{*2}} + \left( \omega^2-V_{\rm D} \right) X = 0 \,,
\eea
where $V_{\rm D} = \omega^2 + {\cal V}$ and
${\cal V}$ is calculated from Eq.~(15) with Eq.~(13) of Ref.~\cite{Detweiler:1977gy}
(or from Eq.~\eqref{eq:U} with Eqs.~\eqref{eq:U0} and \eqref{eq:G} in this paper
where $U$ is identical to ${\cal V}$ and $\gamma'=0$,
see also Eq.~(27) of Ref.~\cite{Seidel:1989bp}
to correct a typo in Eq.~(B23) of Ref.~\cite{Detweiler:1977gy}).
When we adopt $b_2 = - 3 ( a^2 + am/\sigma)$, the potential takes
a slightly simpler expression, 
given in Eq.~(B24) of Ref.~\cite{Detweiler:1977gy} as
\bea
 {\cal V} &=& \frac{-K^2+\Delta \lambda}{(r^2+a^2)^2}
 + \frac{2\Delta(r^3M+a^4)}{r^2(r^2+a^2)^3} + \frac{3a^2\Delta^2}{(r^2+a^2)^4}
 \cr &&
 - \frac{4\lambda \rho^2 \Delta
 \left[-2\lambda \rho^2(r^2-a^2)+2r(rM-a^2)(4\lambda r+6M+\kappa_2) \right]}
 {r^2 (r^2+a^2)^2\left[ 2\lambda r^2+(6M+\kappa_2)r-2\lambda(a^2+am/\sigma) \right]^2} \,.
\eea
As noted in Sec.~\ref{sec:SN},
the above potential $V_{\rm D}$ is the real potential
for the real frequency $\omega=-\sigma$.
However, since the QNM frequencies are complex, $V_{\rm D}$ becomes complex
in our analysis.

\begin{figure}[!ht]
\begin{center}
 \includegraphics[width=0.48\textwidth,clip=true]{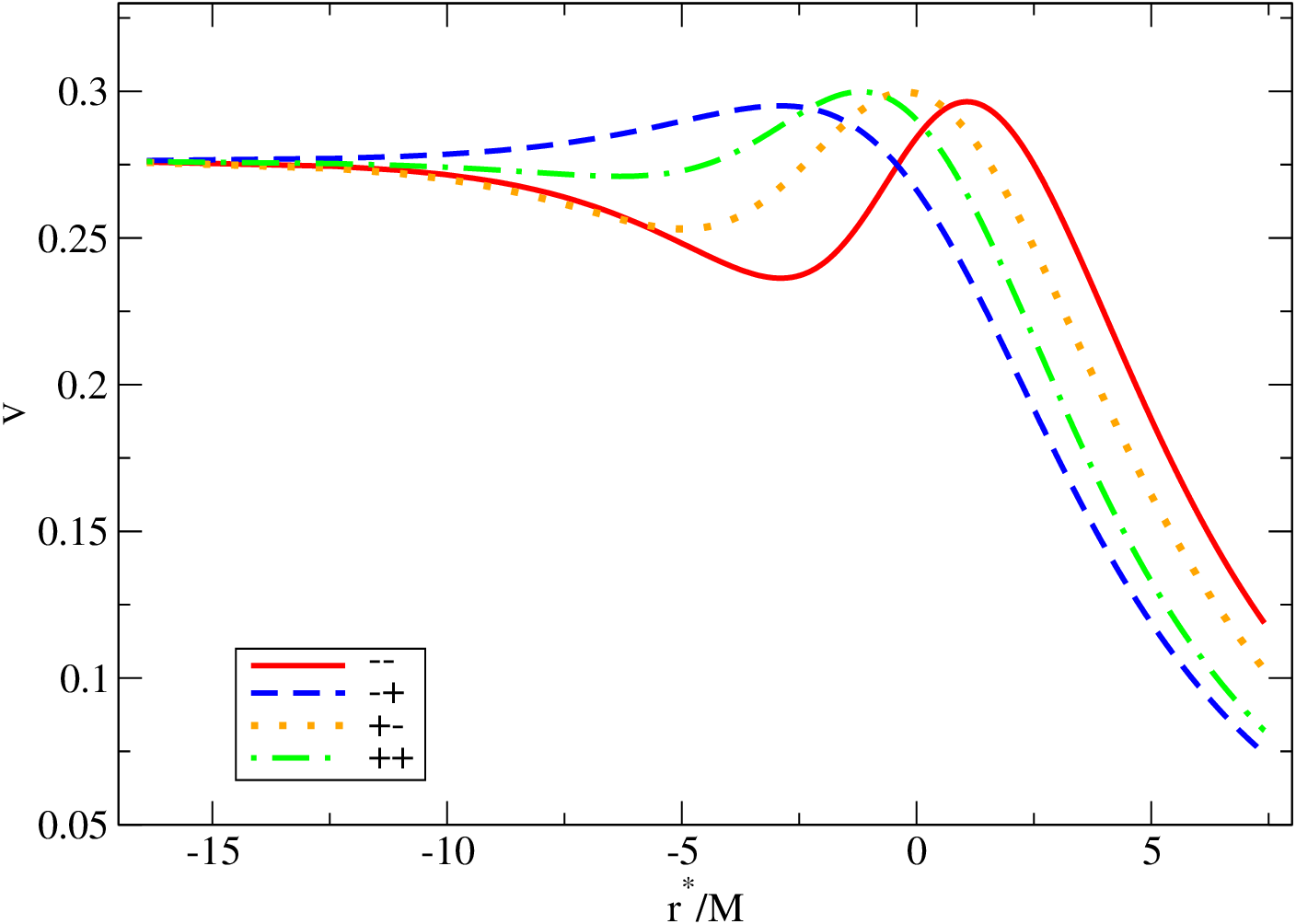}
 \includegraphics[width=0.48\textwidth,clip=true]{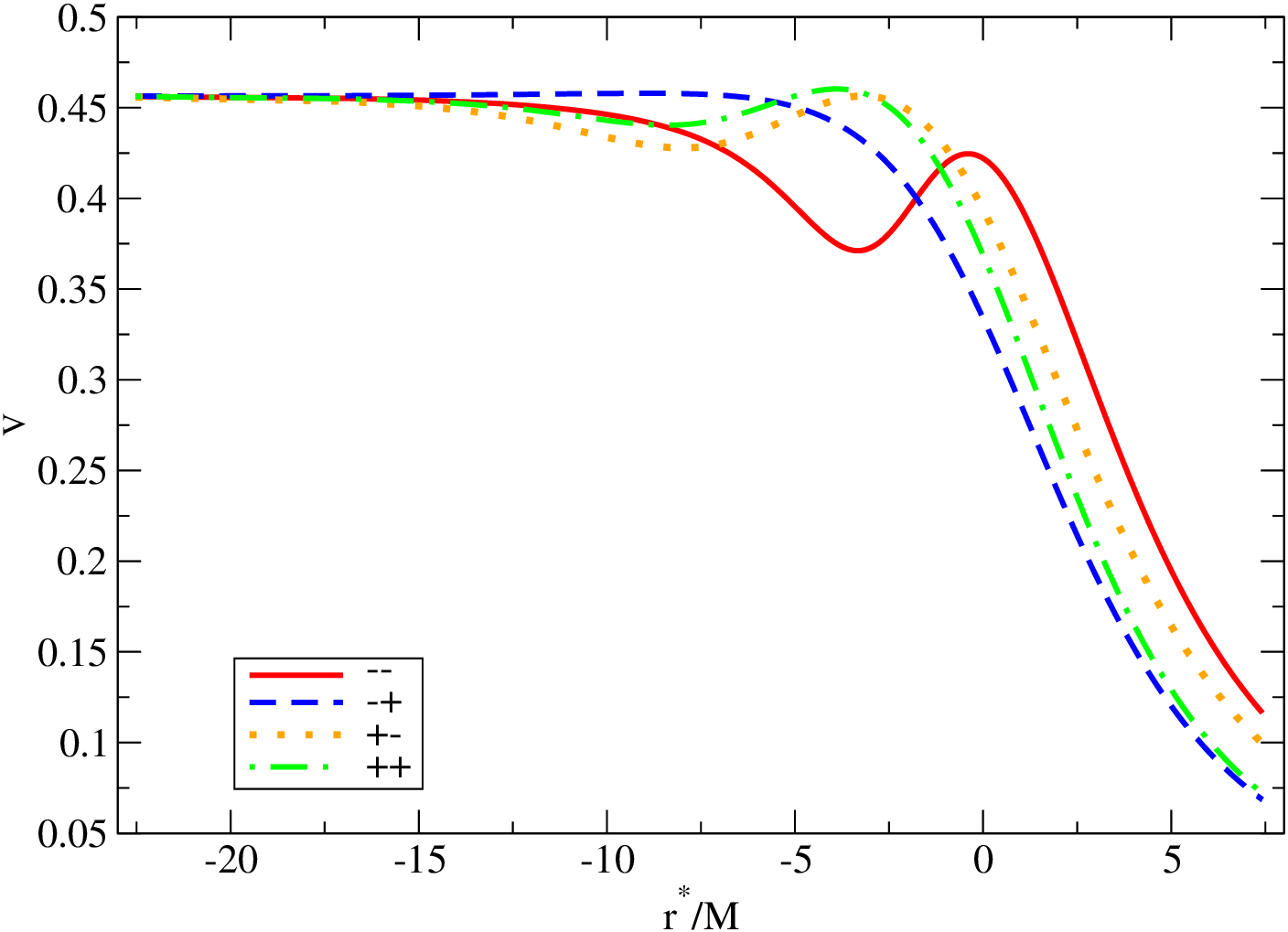}
\end{center}
 \caption{$|V|$ with ($\ell=2,\,m=2)$
 for $a/M=0.7$ and the QNM frequency $M\omega=0.5326-0.0808i$
 (left), and $a/M=0.9$ and $M\omega=0.6716-0.0649i$ (right)
 as a function of $r^*/M$ where we set $M=1$. The first and second
 signatures denote those of $b_2$ in Eq.~\eqref{eq:pmb2} and $\kappa_2$
 in Eq.~\eqref{eq:pmkappa2}, respectively. The ($-+$) case gives $V_{\rm D}$.}
 \label{fig:D_pm0709}
\end{figure}

When we take 
the positive (negative) signature for $\kappa_2$ in Eq.~\eqref{eq:pmkappa2},
the potential $V_{\rm D}$ becomes 
the Zerilli $V_{\rm Z}$ (Regge-Wheeler $V_{\rm RW}$) potential in the Schwarzschild limit.
As shown in Fig.~\ref{fig:D_pm0709} for the $a/M=0.7$ case, the best potential
$V_{\rm D}$ with $(\ell=2,\,m=2$) in order to apply the WKB approximation
is obtained by choosing the negative signature for $b_2$ in Eq.~\eqref{eq:pmb2}
and the positive signature for $\kappa_2$ in Eq.~\eqref{eq:pmkappa2}.
In this case, the extremum is clearly unique. We call this situation the ($-+$) case.
As a reference, we also present the $a/M=0.9$ case in
the right panel of Fig.~\ref{fig:D_pm0709}.

Seidel and Iyer~\cite{Seidel:1989bp}
found that the positive signature of $b_2$ is the best choice
for the $m>0$ mode in the case with the negative signature of $\kappa_2$,
i.e., the ($+-$) case, for example, in Fig.~\ref{fig:D_pm0709}.
In their analysis of Ref.~\cite{Seidel:1989bp}, a higher order WKB approximation was employed
(see also a Kokkotas's work~\cite{Kokkotas:1993ef}).
They expand their real potential $V_{\rm SI}$, i.e., the ($+-$) case
for real frequency $\omega$ and $r_0$ which is the value of $r$
such that $V_{\rm SI}$ is maximized as
\begin{eqnarray}
 V_{\rm SI} &=& V_s+V_1(a\omega)+V_2(a\omega)^2+V_3(a\omega)^3+V_4(a\omega)^4
 + \cdot\cdot\cdot \,, \cr
 r_0 &=& r_s+r_1(a\omega)+r_2(a\omega)^2+r_3(a\omega)^3+r_4(a\omega)^4
 + \cdot\cdot\cdot \,,
\end{eqnarray}
where $V_s$ ($=V_{\rm RW}$) and $r_s$ ($=r_0^{\rm RW}$)
are the potential for the Schwarzschild case and the value of 
$r$ such that $V_s$ is maximized, respectively.
Then, they solve
\begin{equation}
 \frac{dV_{\rm SI}}{dr^*} = 0 \,,
\end{equation}
to determine $r_1,\,r_2,\,r_3,\,r_4,\,\cdot\cdot\cdot$.
To calculate the ($n=0$) QNM frequency, they solve
\begin{equation}
\omega^2-\phi_g(a,\,\omega,\,\ell,\,m) = 0 \,,
\end{equation}
where the function $\phi_g(a,\,\omega,\,\ell,\,m)$ can be derived
by their Eq.~(2).
After obtaining a complex QNM frequency, $r_0$ becomes complex.
This is the reason why we do not adopt their method
since our purpose is not to obtain the accurate QNM frequency
(see e.g., Ref.~\cite{Berti:2009kk} for inaccuracy of the WKB method),
but to establish the physical picture that the QNM brings us information
around the peak radius $r_0^*$.

\begin{figure}[!ht]
\begin{center}
 \includegraphics[width=0.48\textwidth,clip=true]{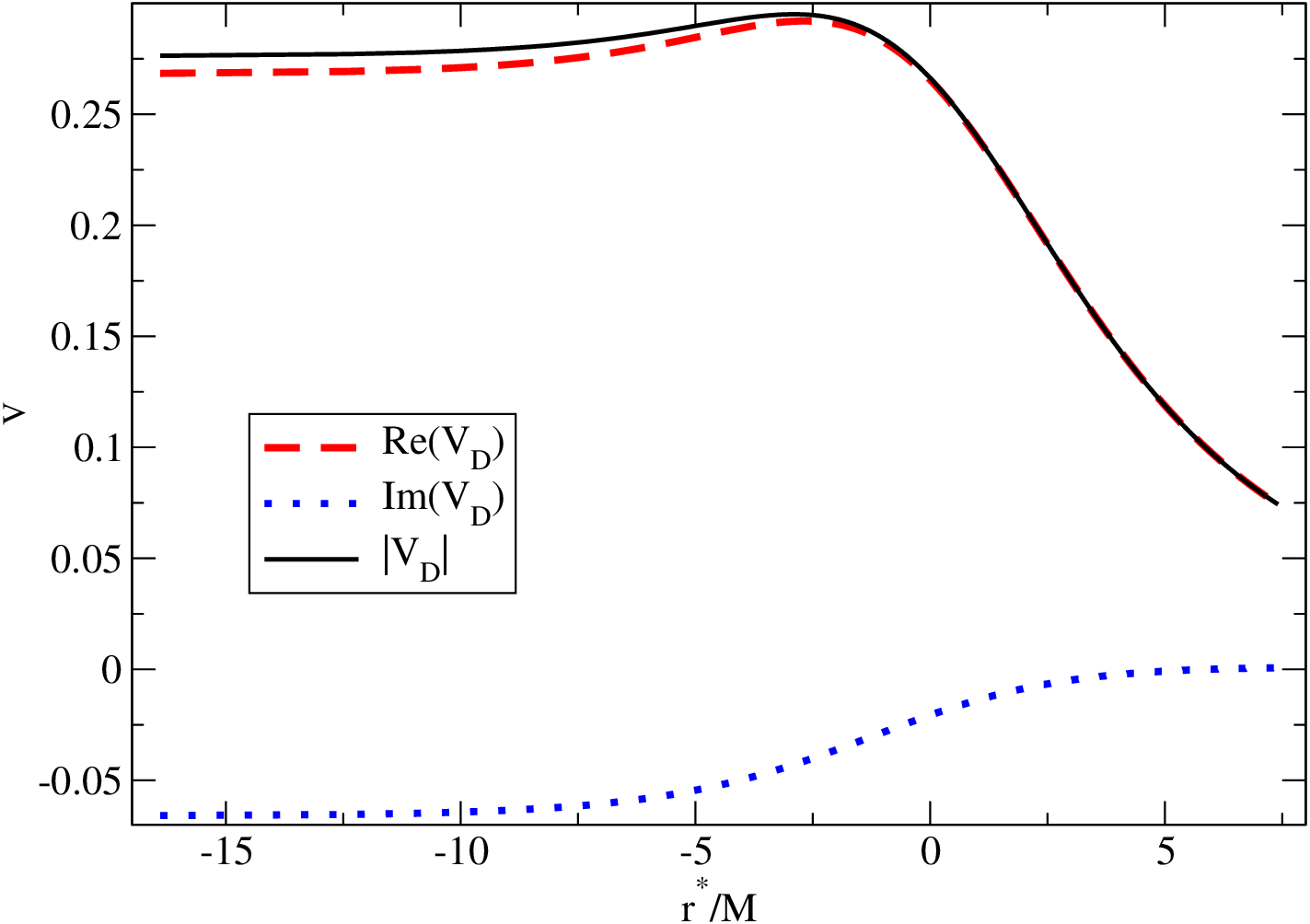}
 \includegraphics[width=0.47\textwidth,clip=true]{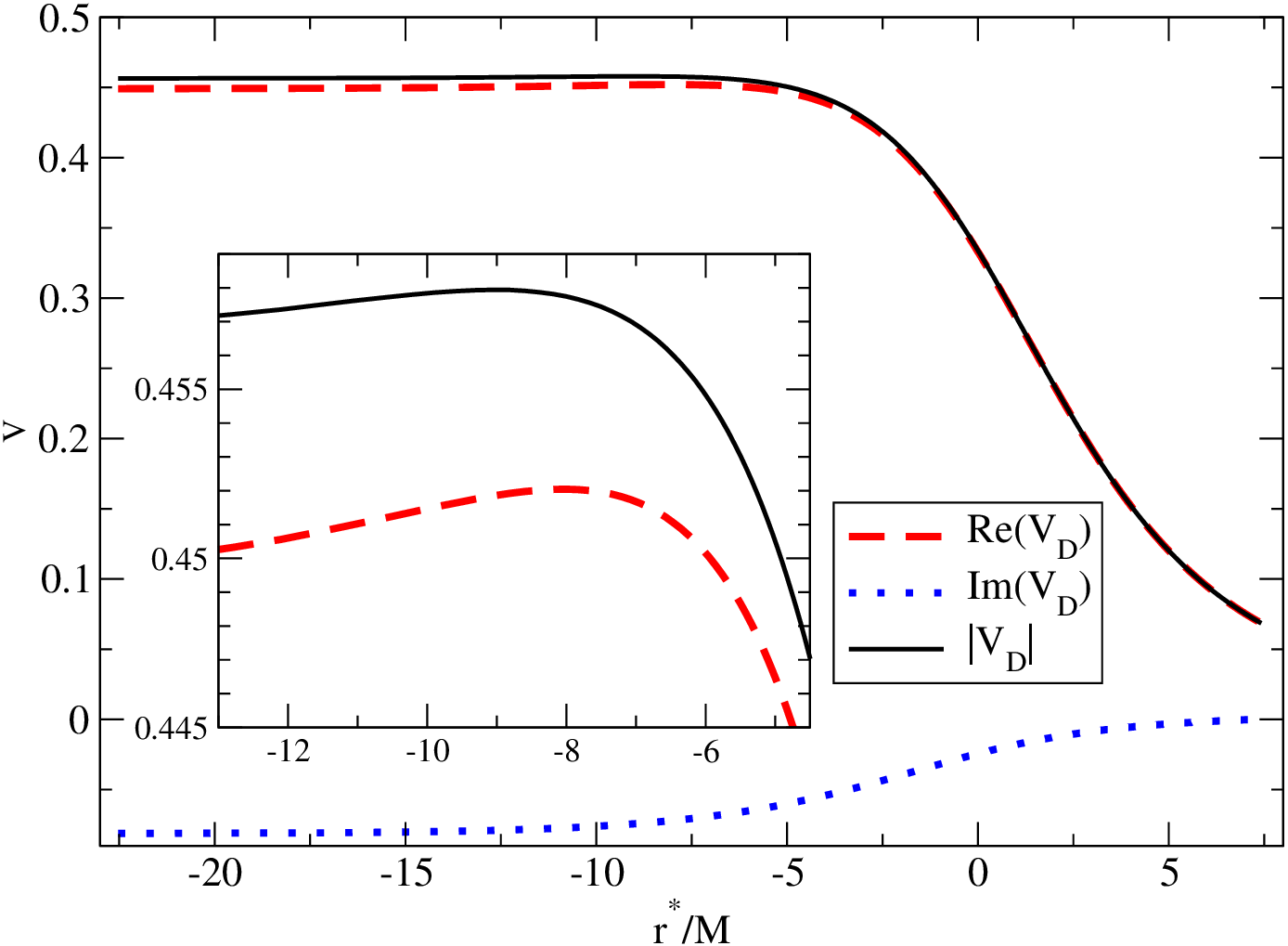}
\end{center}
 \caption{${\rm Re}(V_{\rm D})$, ${\rm Im}(V_{\rm D})$ and $|V_{\rm D}|$ 
 with $(\ell=2,\,m=2$)
 for $a/M=0.7$ and the QNM frequency $M\omega=0.5326-0.0808i$ (left),
 and $a/M=0.9$ and $M\omega=0.6716-0.0649i$ (right)
 as a function of $r^*/M$ where we set $M=1$.
 The contribution of the imaginary part is small.}
 \label{fig:D_mp0709}
\end{figure}

\begin{figure}[!ht]
\begin{center}
 \includegraphics[width=0.48\textwidth,clip=true]{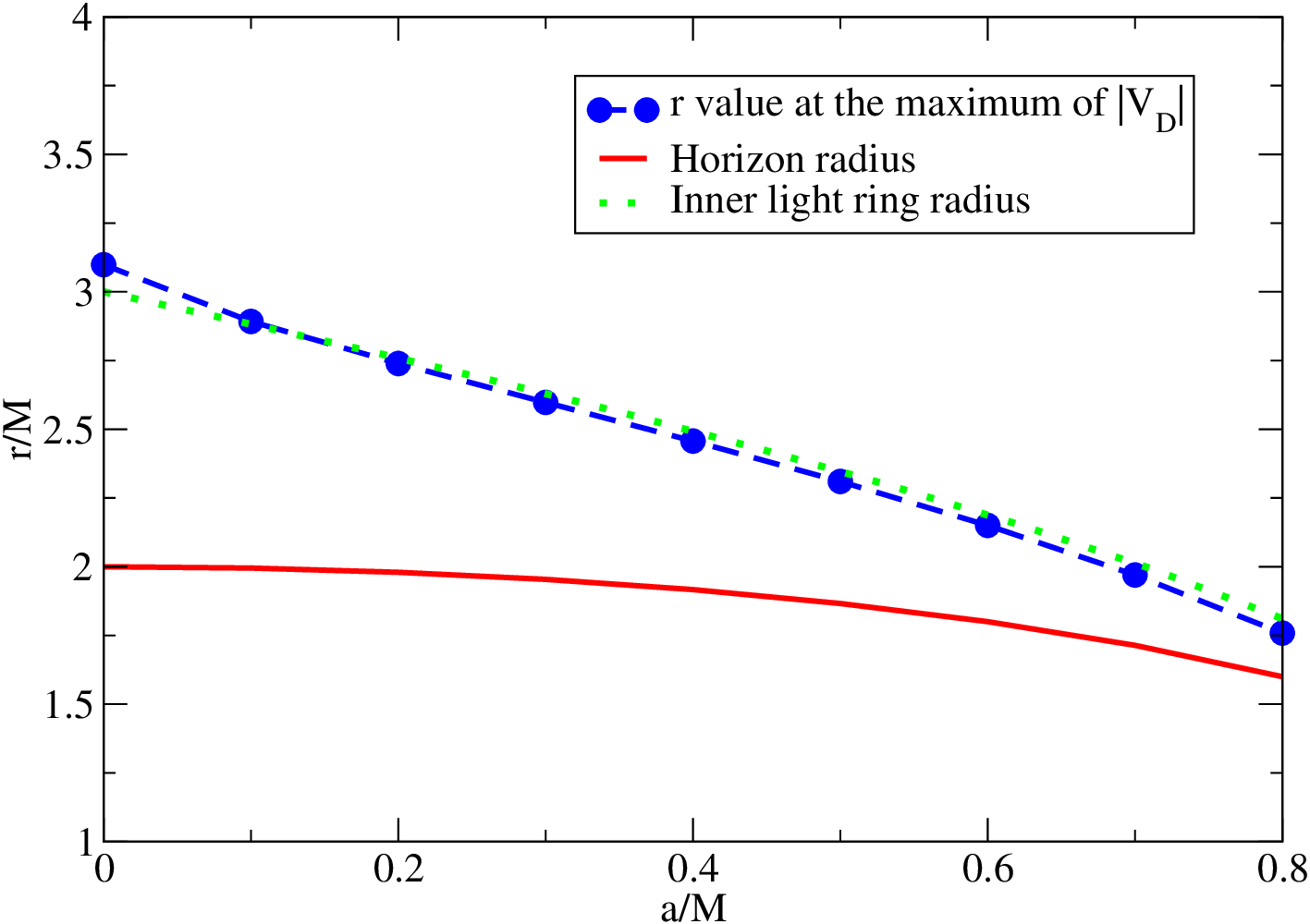}
\end{center}
 \caption{The location of the maximum of the absolute value of the potential
 $|V_{\rm D}|$ with $(\ell=2,\,m=2$),
 the event horizon $r_{+}=M + \sqrt{M^2-a^2}$,
 and the inner light ring radius $r_{\rm lr}=2M(1+\cos((2/3)\cos^{-1}(-a/M)))$
 evaluated for various spin parameters $a$.}
 \label{fig:locationD}
\end{figure}

\begin{figure}[!ht]
\begin{center}
 \includegraphics[width=0.48\textwidth,clip=true]{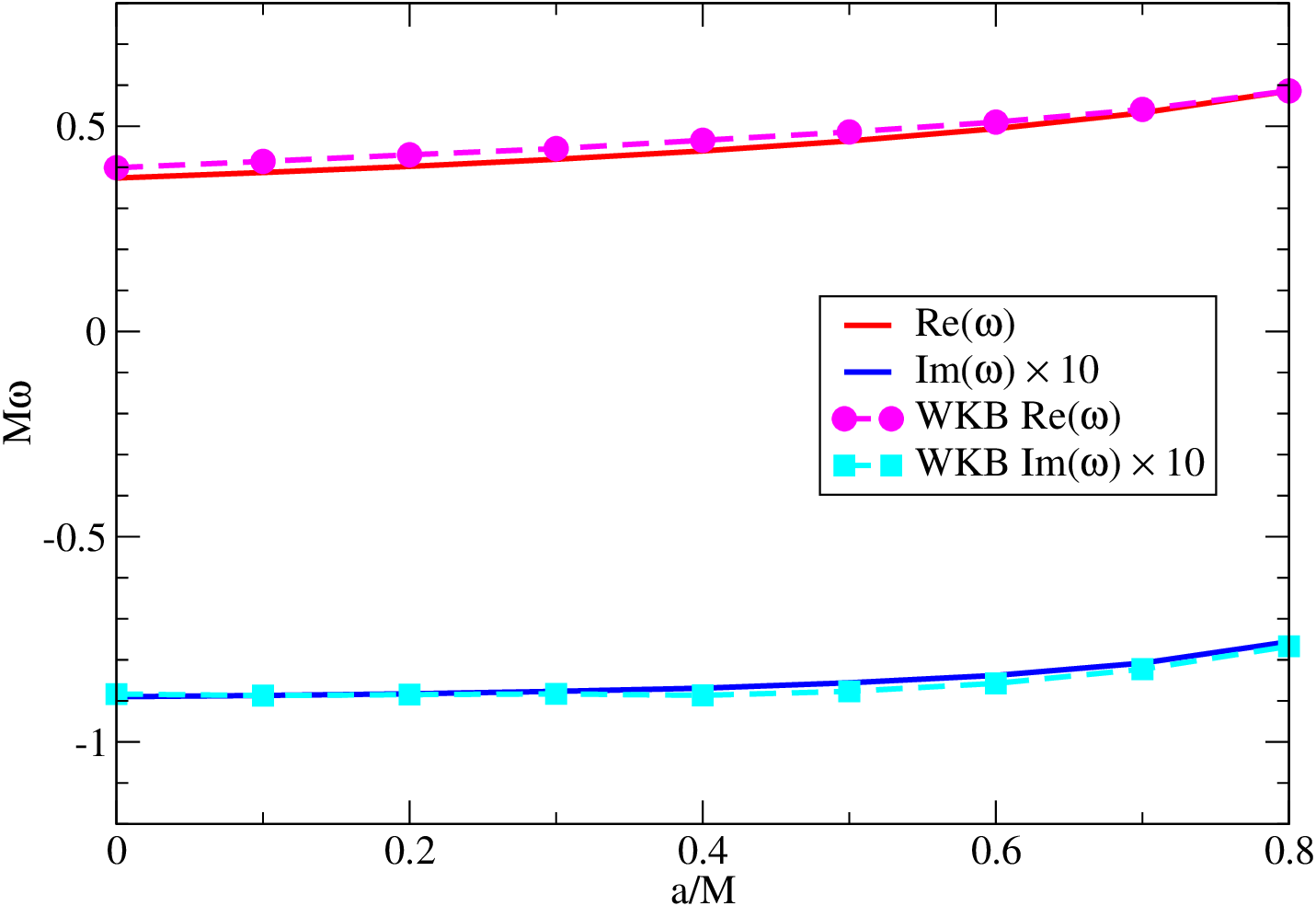}
\end{center}
 \caption{The real and imaginary parts of the fundamental ($n=0$)
 QNM frequencies with $(\ell=2,\,m=2$) evaluated for various spin parameters $a$.
 The exact frequencies ${\rm Re}(\omega)$ and ${\rm Im}(\omega)$
 are from Ref.~\cite{Berti:2005ys}.}
 \label{fig:frequenciesD}
\end{figure}

\begin{figure}[!ht]
\begin{center}
 \includegraphics[width=0.48\textwidth,clip=true]{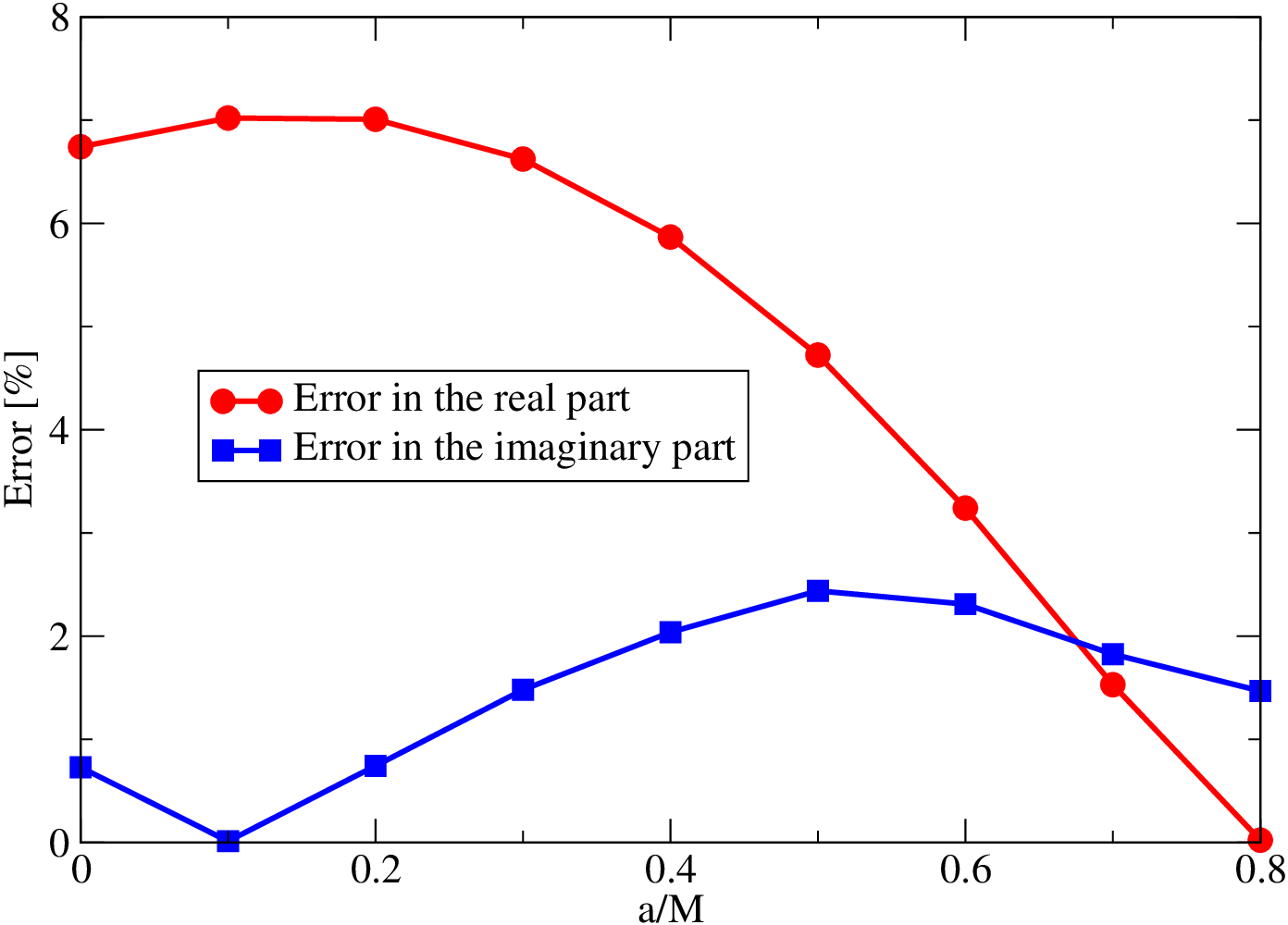}
\end{center}
 \caption{Absolute value of relative errors for the QNM frequencies
 with $(\ell=2,\,m=2$)
 between the exact value and that of the WKB approximation in Fig.~\ref{fig:frequenciesD}.}
 \label{fig:errorsD}
\end{figure}

In our analysis for the ($-+$) case, the $n=0$ QNM frequency is calculated by
\begin{equation}
 (\omega_r+i\omega_i)^2
 = V_{\rm D}(r^*_0)-i\sqrt{-\frac{1}{2}
 \left. \frac{d^2V_{\rm D}}{dr^{*2}} \right|_{r^*=r^*_0}} \,.
 \label{eq:QNM_D}
\end{equation}
Again, this is an approximation because we adopt
the peak of  $|V_{\rm D}|$ as $r^*_0$,
and the approximation will be worse when ${\rm Im}(V_{\rm D})$ becomes large.
The location of the maximum of the absolute value of the potential $|V_{\rm D}|$
with ($\ell=2,\,m=2$)
(see Fig.~\ref{fig:D_mp0709} for the smallness of the imaginary part of the potential),
and the real and imaginary parts of the $n=0$ QNM frequencies
are shown in Figs.~\ref{fig:locationD} and \ref{fig:frequenciesD}, respectively.
Since the contribution of the imaginary part is small,
we use the peak location of $|V_{\rm D}|$ as $r^*_0$ in the WKB approximation.
In Fig.~\ref{fig:errorsD}, we show the errors
in the frequencies. From Fig.~\ref{fig:locationD},
we see that for $a/M=0.7$ the peak of $|V_{\rm D}|$ with $(\ell=2,\,m=2$)
is $r^* \sim -2M$, i.e., $r\sim 2M$,
which is similar to the $V_{\rm SN}$ case in Sec.~\ref{sec:SN}.
Fig.~\ref{fig:errorsD} shows that the error of the QNM frequencies is $< 7\%$,
which is also similar to the $V_{\rm SN}$ case presented in Sec.~\ref{sec:SN}.

%%%%%%%%%%%%%%%%%%%%%%%%%%%%%%%%%%%%%%%
\section{Discussions}\label{sec:dis}
%%%%%%%%%%%%%%%%%%%%%%%%%%%%%%%%%%%%%%%

\begin{figure}[!ht]
\begin{center}
 \includegraphics[width=0.48\textwidth,clip=true]{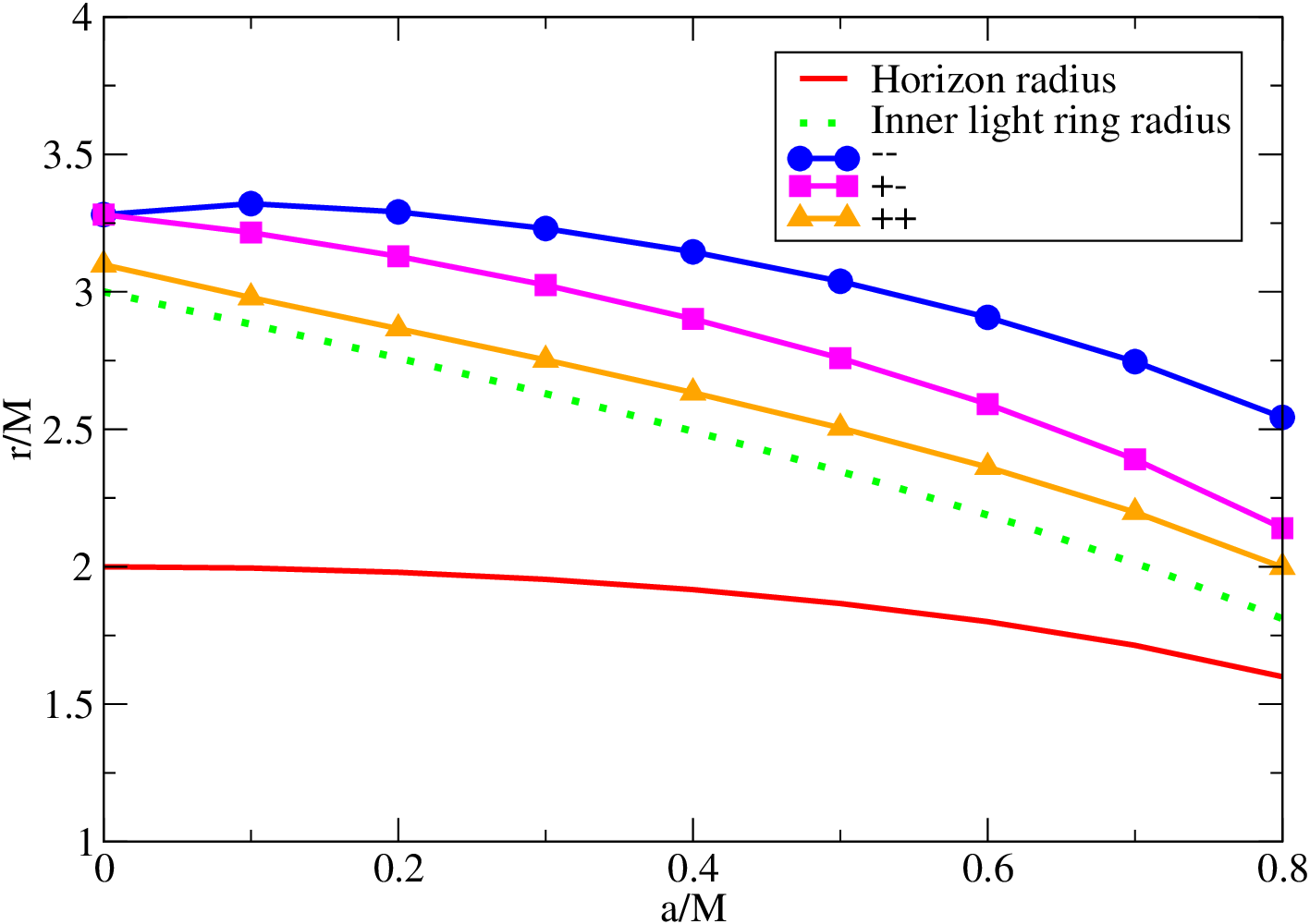}
\end{center}
 \caption{The location of the maximum of the absolute value of the potential
 with $(--)$, $(+-)$ and $(++)$ with $(\ell=2,\,m=2$),
 the event horizon $r_{+}=M + \sqrt{M^2-a^2}$,
 and the inner light ring radius $r_{\rm lr}=2M(1+\cos((2/3)\cos^{-1}(-a/M)))$
 evaluated for various spin parameters $a$.}
 \label{fig:locationDmp}
\end{figure}

\begin{figure}[!ht]
\begin{center}
 \includegraphics[width=0.32\textwidth,clip=true]{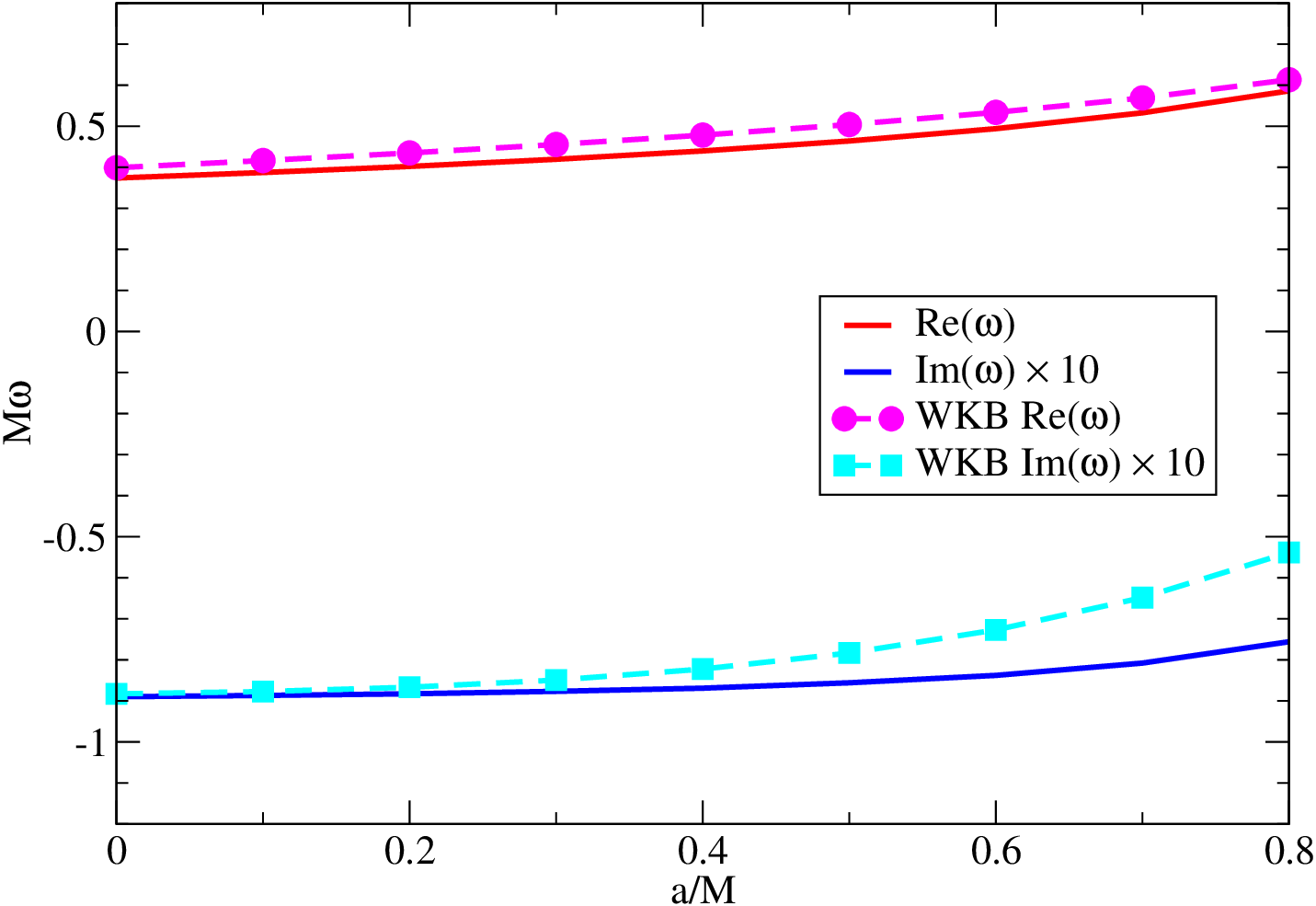}
 \includegraphics[width=0.32\textwidth,clip=true]{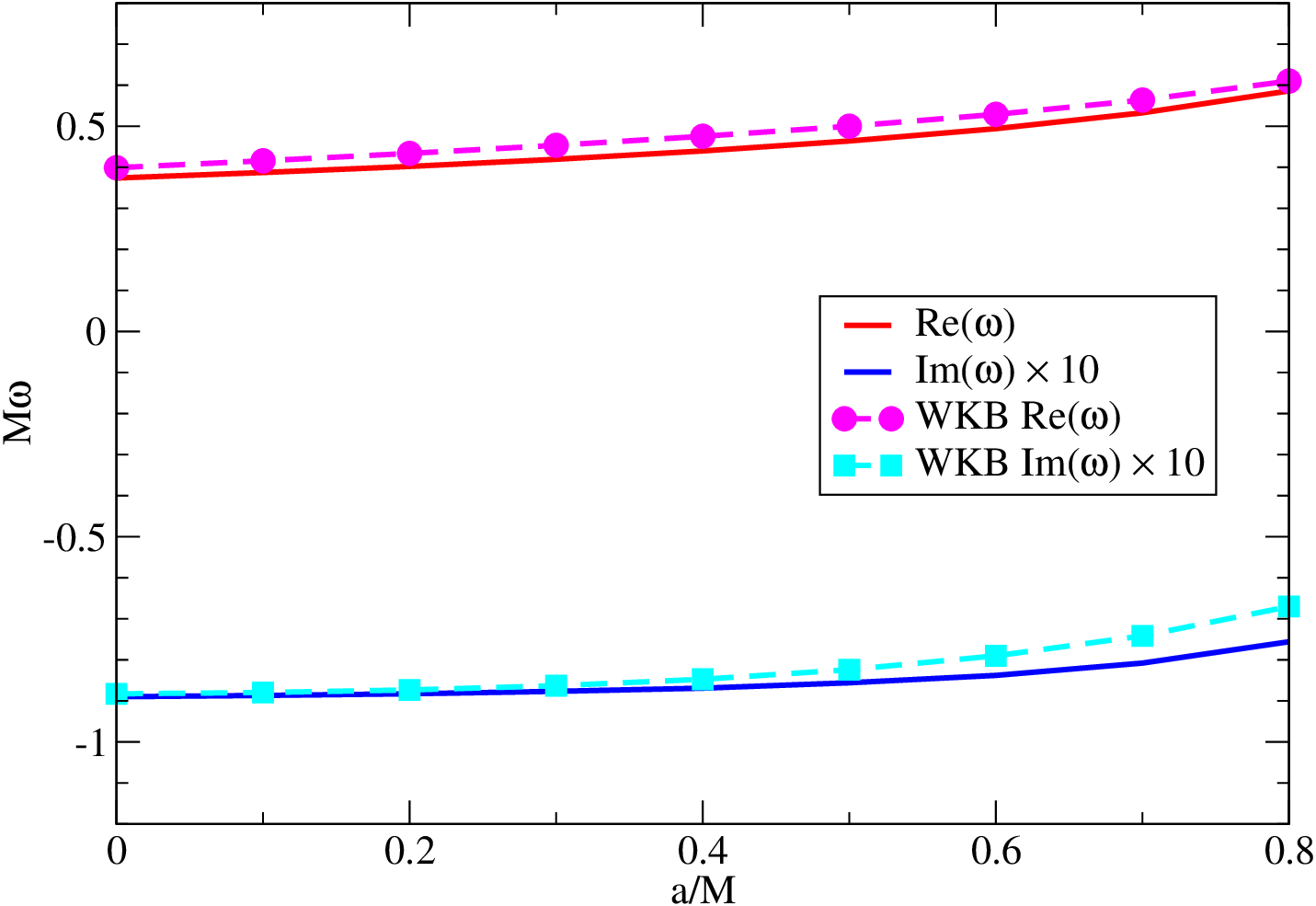}
 \includegraphics[width=0.32\textwidth,clip=true]{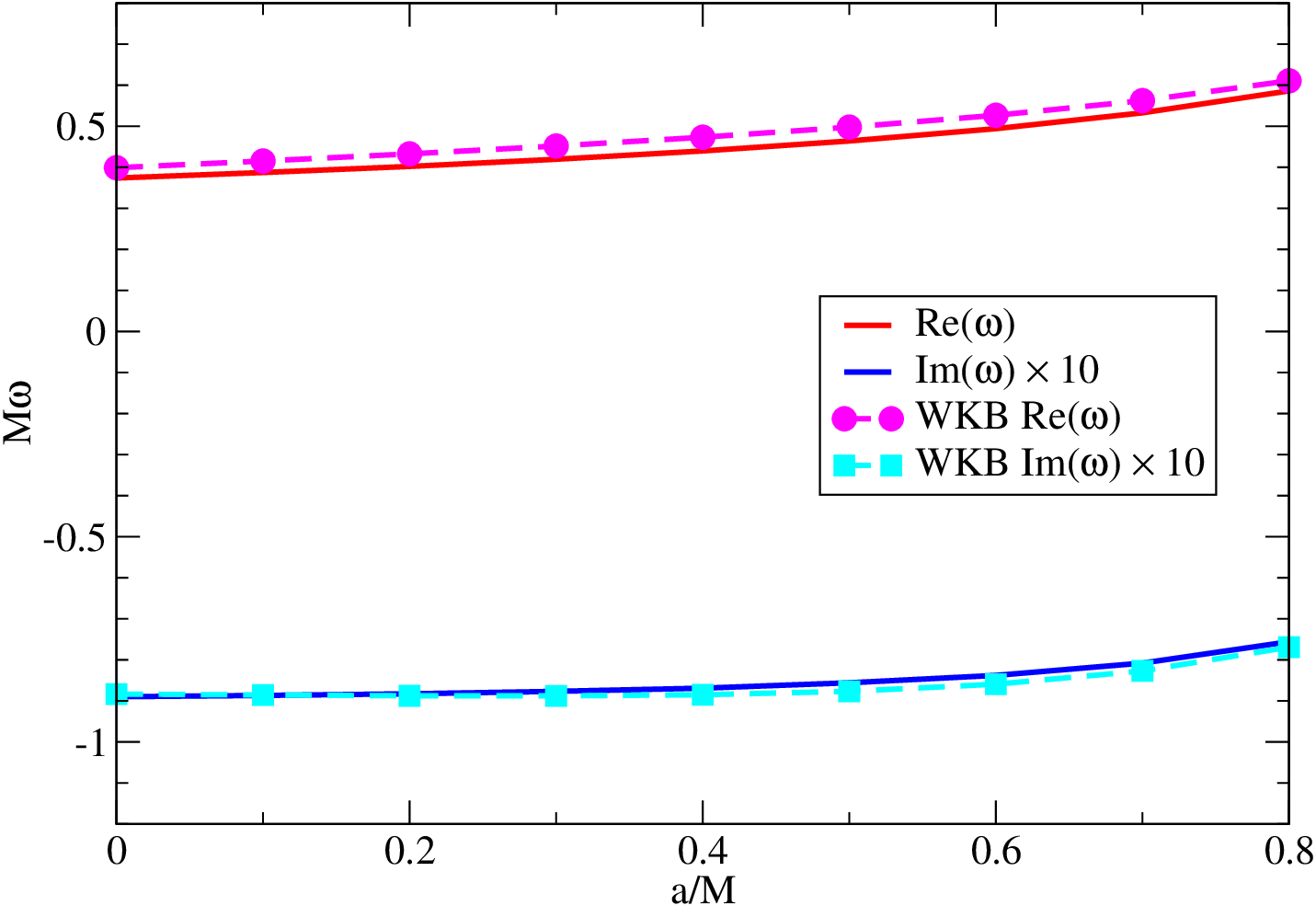}
\end{center}
 \caption{The real and imaginary parts of the fundamental ($n=0$)
 QNM frequencies with $(\ell=2,\,m=2$)
 evaluated for various spin parameters $a$.
 The exact frequencies ${\rm Re}(\omega)$ and ${\rm Im}(\omega)$
 are from Ref.~\cite{Berti:2005ys}.
 The left, center and right panels show
 the ($--$), ($+-$) and ($++$) cases, respectively.}
 \label{fig:frequenciesDpm}
\end{figure}

\begin{figure}[!ht]
\begin{center}
 \includegraphics[width=0.32\textwidth,clip=true]{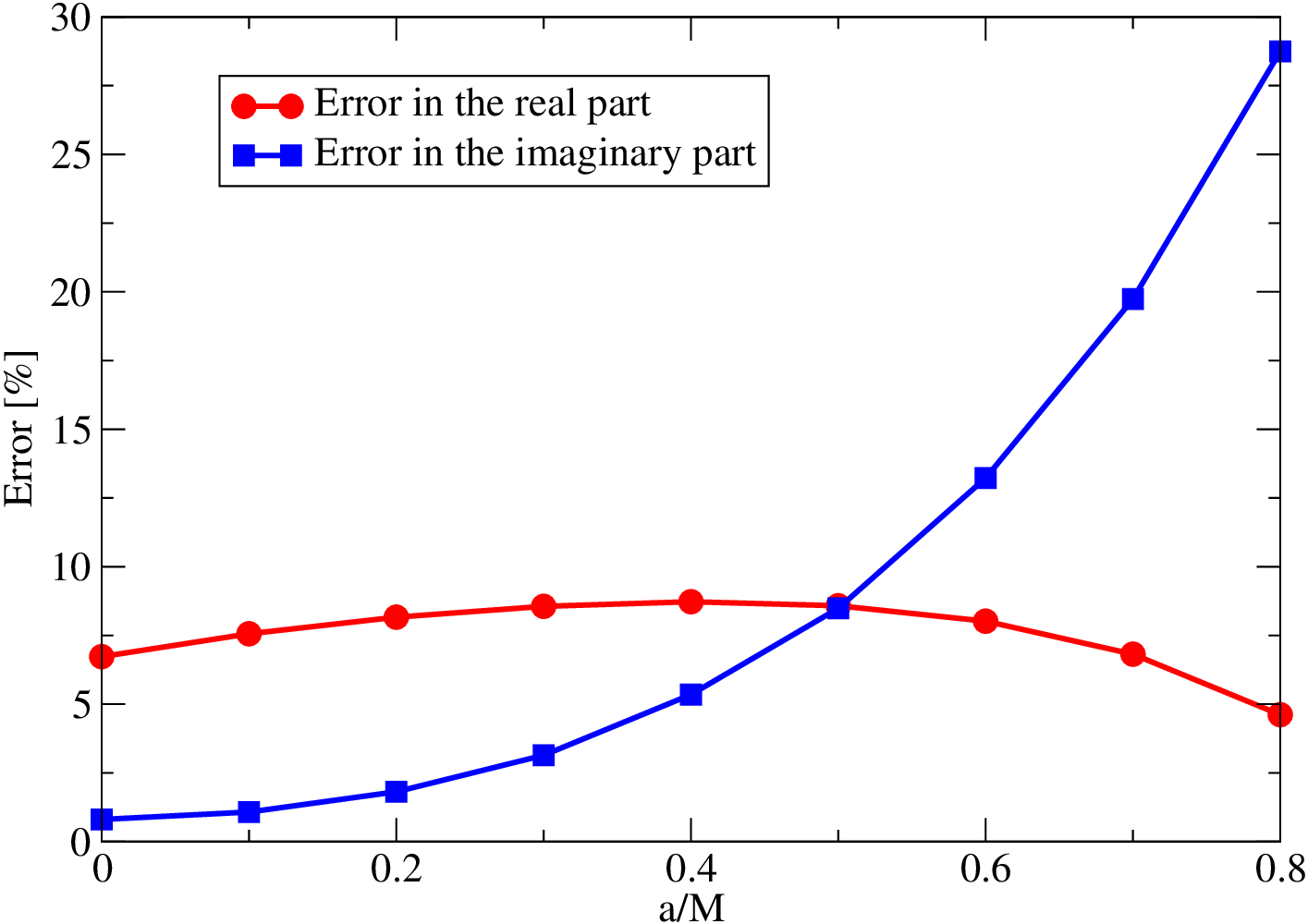}
 \includegraphics[width=0.32\textwidth,clip=true]{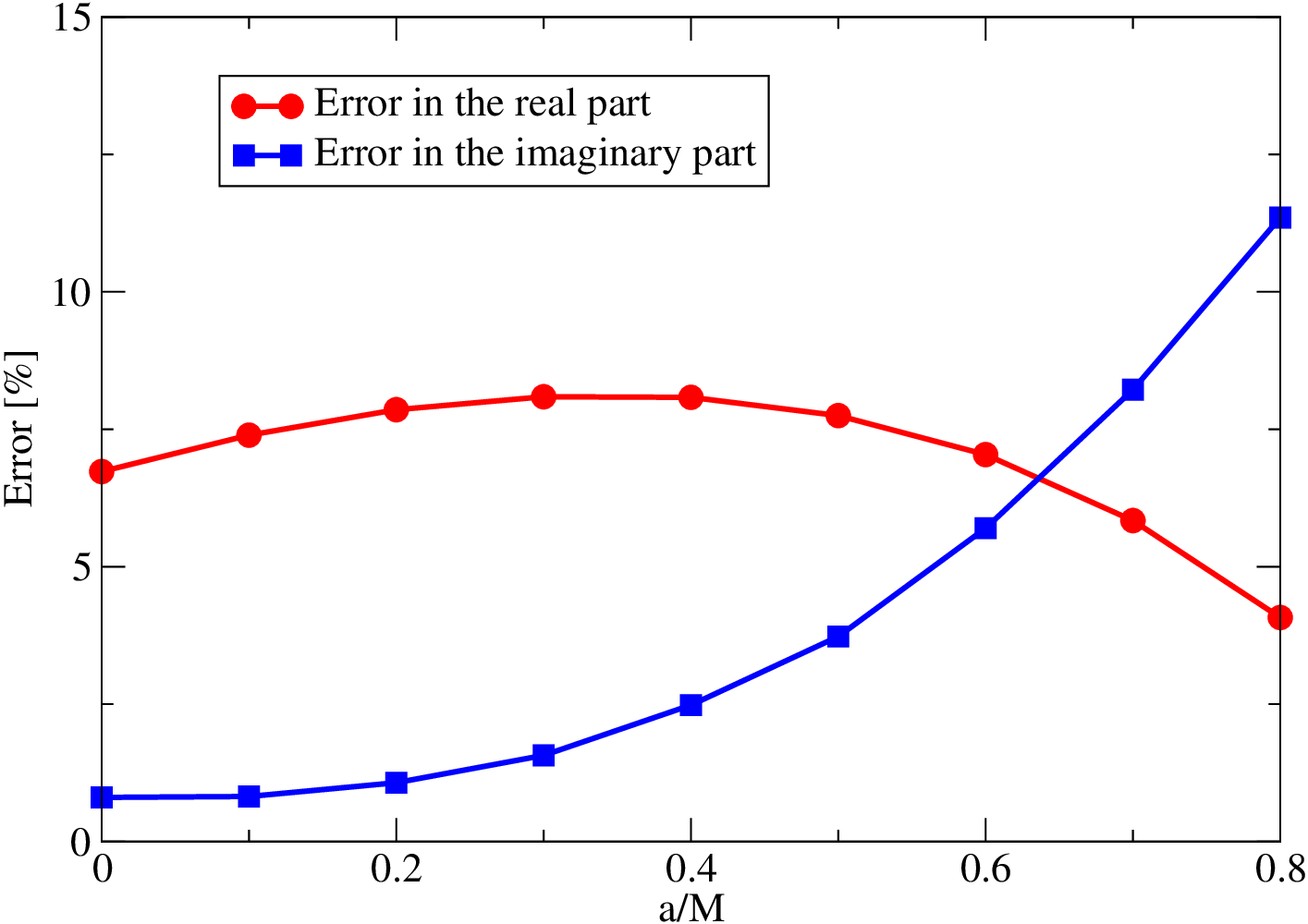}
 \includegraphics[width=0.32\textwidth,clip=true]{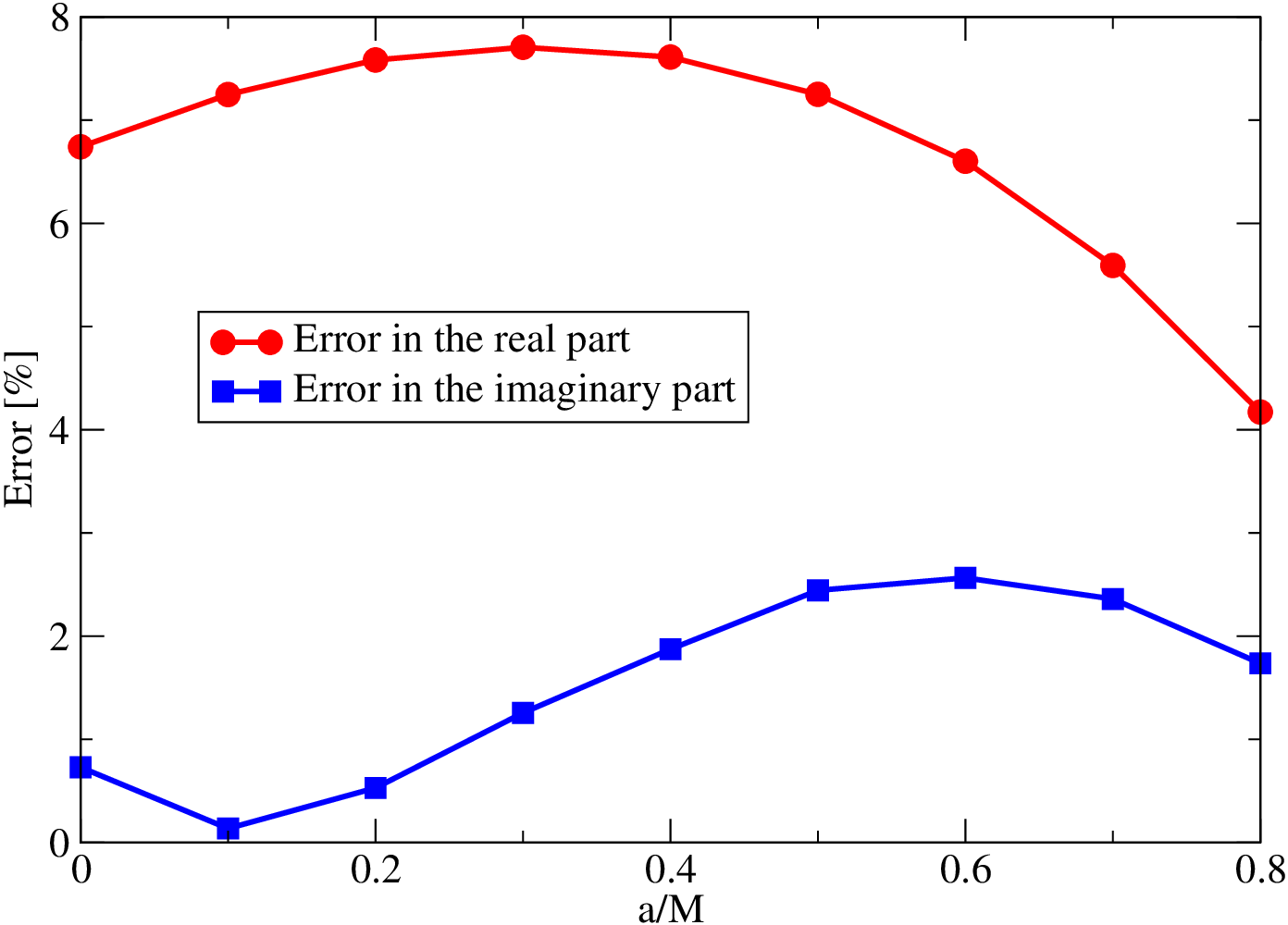}
\end{center}
 \caption{Absolute value of relative errors for the QNM frequencies with $(\ell=2,\,m=2$)
 between the exact value and that of the WKB approximation
 in Fig.~\ref{fig:frequenciesDpm}.
 The left, center and right panels show
 the ($--$), ($+-$) and ($++$) cases, respectively.}
 \label{fig:errorsDpm}
\end{figure}

In the case of the Chandrasekhar and Detweiler equation,
we have used $V_{\rm D}$, i.e, the ($-+$) case.
It might be interesting to see the peak locations and 
the QNM frequencies in the WKB analysis for the other cases.
Fig.~\ref{fig:locationDmp} shows the location of the maximum
of the absolute value of the potential with $(\ell=2,\,m=2$)
for the $(--)$, $(+-)$ and $(++)$ cases.
As seen in Fig.~\ref{fig:D_pm0709}, the peak location depends
on the choice of the signatures in $b_2$ and $\kappa_2$.
We find that the $(--)$, $(+-)$ and $(++)$ cases
do not give a better result for the QNM frequency (see Fig.~\ref{fig:frequenciesDpm})
than the $(-+)$ case
in applying the WKB approximation of Eq.~\eqref{eq:QNM_D}.
The relative errors in the evaluation of the QNM frequencies
compared with the exact value are presented in Fig.~\ref{fig:errorsDpm}.
In particular, the imaginary part of the QNM frequency deviates from the exact one
substantially for $a/M \gtrsim 0.7$.
This is not mainly due to the crude approximation for $r^*_0$ adopted 
in this paper. Even if we use $r^*_0$ corresponding to the complex 
root of $dV/dr_*=0$, 
the errors necessarily grow for a larger value of $a/M$. (Although the 
estimate of the QNM frequency changes, the tendency that the error 
increases in that regime is not altered.) The main cause of 
the error would be due to the appearance of the second root of $dV/dr^*=0$ 
near the real axis of $r_*$, which we can imagine 
for $V_{SN}$, $(--)$, $(+-)$ and $(++)$ cases for $a/M\gtrsim 0.7$ as 
the appearance of double extrema or wavy shape in the plot of $|V|$. 
The appearance of the second root is confirmed also by direct calculation. 
By contrast, $V_{D}$ does not seem to suffer from any significant contribution 
from other roots up to $a/M=0.8$.

Our purpose of the current study is to establish the picture 
that the QNMs are approximately originating from the peak location of 
the potential. The peak location varies depending on the choice of 
the potential, but the variance is not very large as seen in
Figs.~\ref{fig:location}, \ref{fig:locationD} and
\ref{fig:locationDmp}. Furthermore, 
we find that the estimate of the QNM frequencies by using 
$V_{\rm SN}$, $V_{\rm D}$ and $(++)$ cases 
is better than that in the $(--)$ and $(+-)$ cases. 
If the cases that give worse estimate of the QNM frequencies, i.e., the
$(--)$ and $(+-)$ cases are excluded, the variance of the peak location 
becomes even smaller.

In the high frequency regime the real and imaginary parts of 
the QNM frequencies are thought to be 
related to the orbital frequency of the light ring orbit and the 
Lyapunov exponent of the perturbation around it, 
respectively (see Refs.~\cite{Cardoso:2008bp,Dolan:2009nk,Dolan:2010wr,
Yang:2012he,Yang:2012pj,Yang:2013uba}
for the related works and a useful lecture note~\cite{Berti:2014bla}).  
In Figs.~\ref{fig:location}, \ref{fig:locationD} and
\ref{fig:locationDmp} we display the inner light ring radius 
on the equatorial plane as a 
reference. From this comparison, the peak location is found to be 
well approximated by the inner light ring radius, especially for 
$V_{\rm SN}$, $V_{\rm D}$ and $(++)$ cases. This fact suggests 
that even for the lowest order QNM with $n=0$ the light ring 
radius is a good approximation for the region that is 
responsible for generation of QNMs (see pioneer works in
Refs.~\cite{Press:1971wr,Goebel:1972}).

One may think that QNM frequencies depend on the boundary 
condition imposed in the very vicinity of the event horizon, and 
therefore modification to general relativity 
through the change of the boundary condition may lead to 
some observable effect. However, we find it unlikely 
to find the effect as modified QNM frequencies, from the following 
consideration. 
Suppose that complete reflection boundary condition is 
imposed at a distance $\delta$ in 
the proper distance from the event horizon. This radius is, roughly speaking,
$M\ln(\delta/M)$, 
in the $r^*$ coordinate, and hence 
it takes time of $O(M\ln(\delta/M))$ 
before a wave completes a round trip between $r^*\sim 2M$ 
and this hypothetical reflection boundary. If we choose the 
Planck length for $\delta$ as expected by 
the stringy modification such as the firewall, 
this time scale for the round trip becomes as long as $\sim 200M$, 
which is much longer than the time scale of the decay of QNMs, 
which is typically $\sim 20M$. 
Therefore, we find that the effect of the modified boundary condition
will not appear as a measurable shift of the QNM frequencies
in near future, even if we assume an extreme situation such
as a complete reflection. However, the effect of the modification of
the boundary condition may arise in a different manner
or in a very accurate future measurement of the QNM frequencies
as an observable signature (see Refs.~\cite{Damour:2007ap,Barausse:2014tra}).
We may come back to this issue in our future publication.

In Ref.~\cite{Nakano:2015uja} 
we discussed binaries such that form  a $60 M_\odot$ BH with $a/M=0.7$.
In the expected noise curve of KAGRA 
[bKAGRA, VRSE(D) configuration] presented in Ref.~\cite{bKAGRA},
the parameter estimation (1$\sigma$) errors for the real and imaginary parts
with SNR $=35$
are $\pm 0.8\%$ and $(-6\%,\,+7\%)$, respectively, by using the Fisher analysis
(see, e.g., Refs.~\cite{Nakano:2003ma,Nakano:2004ib,Tsunesada:2004ft,Tsunesada:2005fe}
for the QNM gravitational wave data analysis).
Therefore, we concluded that we have a good chance to confirm the 
existence of black hole formed as a result of binary merger. 

When we have at hand 
not just a feature of black hole candidates but 
a clear evidence for the existence of BHs by the QNM observation, 
we will have to wait sitting on a gold mine, 
if we cannot find a clue to carry out further research. 
The breakthrough might be brought by the detection of 
the electro-magnetic counterpart.
Let us consider a merged BH with mass $M\sim 60 M_\odot$
moving with the velocity $V \sim {\rm km\,s^{-1}}$ in the interstellar gas
with the number density $n$.
Then, the accretion rate $\dot{M}$ is estimated as
\begin{eqnarray}
 \dot{M} &=& \pi \left(\frac{2M}{V^2}\right)^2 m_p \,n \,V \cr
 &=& 1.34 \times 10^{20}~{\rm g~s^{-1}}
 \left(\frac{M}{60M_\odot} \right)^2 \left(\frac{n}{10^2~{\rm cm^{-3}}}\right)
 \left(\frac{V}{{\rm 1~km~s^{-1}}}\right)^{-3} \cr
 &=& 1.25\times 10^{40}~{\rm erg~ s^{-1}}
 \left(\frac{M}{60M_\odot}\right)^2\left(\frac{n}{10^2~{\rm cm^{-3}}}\right)
 \left(\frac{V}{{\rm 1~km~s^{-1}}}\right)^{-3} \,,
\end{eqnarray}
where $m_p$ is the mass of the proton and $10.4\%$ efficiency of the available gravitational energy
at the innermost stable circular orbit (ISCO) for $a/M=0.7$ is assumed. 
The luminosity is comparable to the Eddington luminosity of $60 M_\odot$ star,
i.e., ${\rm 7.4 \times 10^{39}~erg~s^{-1}}$.
If this energy is emitted mainly in the X-ray,
the wide field X-ray telescope such as ISS lobster can observe  
up to about $\sim 1$~Mpc since the limiting flux is
$\sim {\rm 10^{-10}~erg~cm^{-2}~s^{-1}}$~\cite{Kisaka:2015mza}.
While if the energy is emitted mainly in the optical band,
it can be detected up to $\sim 300$~Mpc by Subaru-HSC~\cite{Subaru-HSC}
and LSST~\cite{LSST} with the limiting magnitude 26 for example. 
It should be noted that, 
since the sky location of this gravitational wave source
is $\sim 1.7^\circ \times 1.7^\circ$ for the event with SNR$=35$, 
this area can be covered by one or a few shots with Subaru-HSC~\cite{Subaru-HSC}
and LSST~\cite{LSST}.

The mass formula of the Kerr BH with the gravitational mass $M$
and the angular momentum $J$ is given by
\begin{equation}
 M^2=M_{\rm irr}^2 + \frac{J^2}{4M_{\rm irr}^2} \,,
\end{equation}
where $M_{irr}$ is the irreducible mass of the Kerr BH
(see, e.g., Ref.~\cite{MTW:1973}). 
Expressing $J = aM = qM^2$, we have
\begin{equation}
 \label{eq:mass}
 M^2=M_{\rm irr}^2+\frac{M^4 q^2}{4M_{\rm irr}^2} \,,
\end{equation}
where $q^2 = (a/M)^2 < 1$.
Eq.~\eqref{eq:mass} is solved as 
\begin{equation}
 M^2 = \frac{2M_{\rm irr}^2}{1+\sqrt{1-q^2}} \,.
\end{equation}
Therefore, the maximum energy available
by the extraction of all the angular momentum is given by
\begin{equation}
 \Delta E = M \left( 1-\sqrt{\frac{1+\sqrt{1-q^2}}{2}} \right) \,.
\end{equation}
For $q=0.7$, $\Delta E$ becomes
\begin{equation}
 \label{eq:del_e}
 \Delta E = 8\times 10^{54}~{\rm erg} \ \left( \frac{M}{60 M_\odot} \right) \,.
\end{equation}
This energy is so huge that the possible electromagnetic counter part of
Pop III $30M_\odot$--$30M_\odot$ mergers exists.
Therefore, one might think that the Blandford-Znajek (BZ)~\cite{Blandford:1977ds}
luminosity might be much larger. Now the rotational velocity at ISCO
is $v_{\rm ISCO} \sim 10^{10}~{\rm cm~s^{-1}}$.
Assuming the equipartition of the ram pressure energy
 to the magnetic field energy at ISCO, the magnetic strength $B$ is roughly estimated as 
 \begin{eqnarray}
 B &\sim& \frac{\sqrt{2\dot{M}v_{\rm ISCO}}}{r_{\rm ISCO}} \cr
 &\sim& 5\times 10^7~{\rm gauss} \,.
\end{eqnarray}
According to Eq.~(A6) of Penna et al.~\cite{Penna:2013rga}, the BZ power
derived from the magnetic flux $\Phi=4\pi B M r_+$ of their Eq.~(A4)
and the angular velocity of the event horizon $\omega_+=q/(2 r_+)$
as
\begin{eqnarray}
 L^{\rm BZ} &\sim& \frac{\pi}{6} \left(\frac{2 GM q}{c^2}\right)^2 c B^2 \cr
 &\sim& 7.4 \times 10^{38}~{\rm erg~s^{-1}} \,,
\end{eqnarray}
which is smaller than the accretion energy unfortunately.
Here, we recovered $c$ and $G$ in the above equation
for convenience.
We need a certain process to increase the magnetic field strength
near the event horizon to liberate the huge energy shown
in Eq.~\eqref{eq:del_e} to be visible in electromagnetic radiation
to identify the position as well as the strong gravity space-time
near the event horizon of the BH
since the mean distance to Pop III BH merger is
$z \sim 0.3$, i.e., $\sim 1.4$~Gpc in the luminosity distance.

Finally we like to mention that the current scenario that SGRBs are identified
with NS--NS and/or NS--BH mergers does not have any smoking gun so far.
For example, it is usually claimed that no SGRB accompanied
the supernova explosion, and hence SGRBs are
different from Long GRBs (LGRBs). However, there are at least
two LGRBs without the supernova association~\cite{Fynbo:2006mw}.
Only the detection of gravitational waves from SGRBs can be the smoking gun
of this scenario.
Concerning this, before 1997 almost all GRB scientists except
for a few people including Paczynski believed that GRBs are not at a cosmological distance
but within at most in the halo of our galaxy from various suggestions,
without any smoking gun.
The cosmological redshift found in the afterglow of GRB 970508 was the smoking gun
of the cosmological origin of GRBs to force more than 2000 papers useless. 
As one of the counter theories against  the current scenario,
there is, for example, an unified model of LGRB, SGRB,
XRR (X-ray Rich GRB) and XRF (X-Ray Flash)
(see, e.g., Yamazaki, Ioka and Nakamura~\cite{Yamazaki:2004ha}).
Even if the current scenario of SGRBs turns out to be incorrect unexpectedly,
Pop III BH--BH binary mergers discussed in this paper would be
another plausible candidates of the important sources of gravitational waves
to search the strong gravity space-time near the event horizon
of the BH of mass $\sim 60 M_\odot$.

%%%%%%%%%%%%%%%%%%%%%%%%%%%%%%%%%%%%%%%
\acknowledgments
%%%%%%%%%%%%%%%%%%%%%%%%%%%%%%%%%%%%%%%

We thank T.~Sakamoto for a careful reading of this
manuscript, and for helpful comments.
We would also like to appreciate the anonymous referee for valuable comments.
This work was supported by MEXT Grant-in-Aid for Scientific Research
on Innovative Areas,
``New Developments in Astrophysics Through Multi-Messenger Observations
of Gravitational Wave Sources'', No.~24103006 (TN, HN, TT) and
by the Grant-in-Aid from the Ministry of Education, Culture, Sports,
Science and Technology (MEXT) of Japan No.~15H02087 (TN, TT).

%%%%%%%%%%%%%%%%%%%%%%%%%%%%%%%%%%%%%%%%
\appendix

%%%%%%%%%%%%%%%%%%%%%%%%%%%%%%%%%%%%%%%
\section{Another estimation}\label{app:ano}
%%%%%%%%%%%%%%%%%%%%%%%%%%%%%%%%%%%%%%%

\begin{figure}[!ht]
\begin{center}
 \includegraphics[width=0.48\textwidth,clip=true]{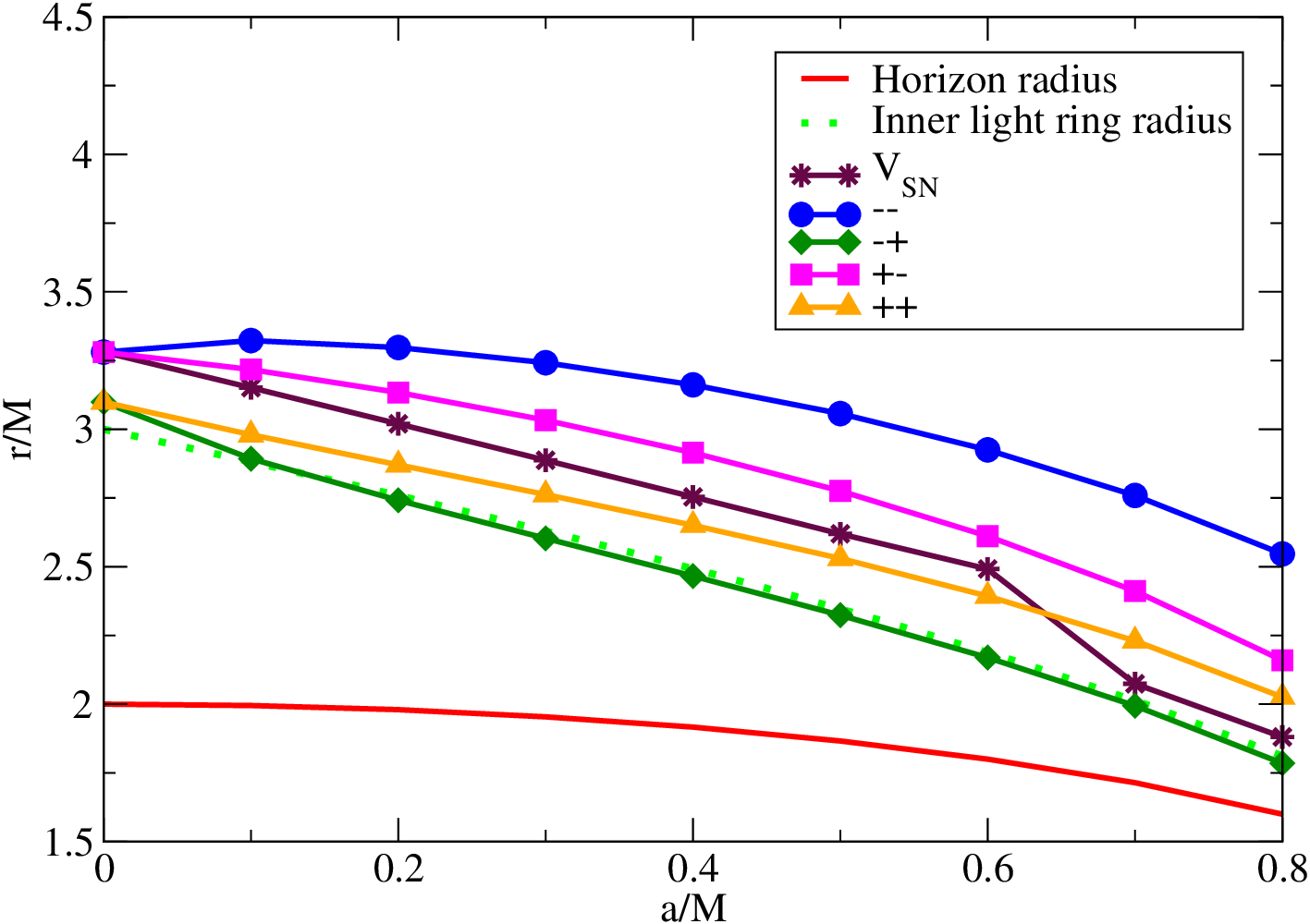}
\end{center}
 \caption{The real part of the radius derived from $dV/dr^*=0$
 in the complex plane.
 Here, the potential $V$ is the Sasak-Nakamura's $V_{\rm SN}$,
 the Detweiler's $(--)$, $(-+)$, $(+-)$ or $(++)$ case with $(\ell=2,\,m=2$).
 We also show the event horizon $r_{+}=M + \sqrt{M^2-a^2}$,
 and the inner light ring radius $r_{\rm lr}=2M(1+\cos((2/3)\cos^{-1}(-a/M)))$
 evaluated for various spin parameters $a$.}
 \label{fig:locationC}
\end{figure}

In this appendix, we summarize the effective peak radius
derived from the real part of the radius which satisfies $dV/dr^*=0$.
Fig.~\ref{fig:locationC} shows the radii obtained from various potentials.
Although it is difficult to understand the jump
in the radius between $a/M=0.6$ and $0.7$
in the analysis the maximum of the absolute value of the potential $|V_{\rm SN}|$,
we find the reason from the calculation of $dV/dr^*=0$ in the complex plane.
For $a/M=0.6$, we have a solution $r_0 = 2.49187 - 0.00352404i$.
On the other hand, we have two solutions for $a/M=0.7$, 
$r_0 = 2.58852 - 0.0588482i$ and $2.07427 + 0.0268525i$.
Here, the peak location of $|V_{\rm SN}|$ for $a/M=0.7$
corresponds to the latter solution.
It is also noted that the imaginary part of the solution for $dV/dr^*=0$
is always an order of magnitude smaller than the real part
in the current analysis between $a/M=0$ and $0.8$,
and we find a good agreement between the peak location of $|V_{\rm SN}|$
and the real part of $r^*_0$ of $dV_{\rm SN}/dr^*=0$,
comparing Figs.~\ref{fig:location}, \ref{fig:locationD} and \ref{fig:locationDmp}
with Fig.~\ref{fig:locationC}.

%%%%%%%%%%%%%%%%%%%%%%%%%%%%%%%%%%%%%%%%

%%%%%%%%%%%%%%%%%%%%%%%%%%%%%%%%%%%%%%%%
\end{document}